\documentclass[12pt]{article}
\usepackage{graphicx, here, amsfonts, epsfig}
\def\Journal#1#2#3#4{{#1} {\bf #2}, #3 (#4)}

\def\YAD{\em Yad. Fiz.}
\def\NCA{\em Nuovo Cimento}

\def\NPB{{\em Nucl. Phys.} B}
\def\PLB{{\em Phys. Lett.}  B}

\def\PRD{{\em Phys. Rev.} D}
\def\ZPC{{\em Z. Phys.} C}
\def\EPG{{\em Eur.Phys.J.} C}

\def\SJP{{\em Sov. Phys. JETP}}

\def\be{\begin{equation}}
\def\ee{\end{equation}}
\def\bea{\begin{eqnarray}}
\def\eea{\end{eqnarray}}
\def \Pom {{\hspace{ -0.05em}I\hspace{-0.25em}P}}
\def \Reg {{\hspace{ -0.05em}I\hspace{-0.25em}R}}

\begin{document}
\begin{center}
{\Large \bf A universal Regge pole model for all vector
meson exclusive photoproduction by real
and virtual photons}

\vskip 1.cm
{E. MARTYNOV$^{a,}$\footnote{\it E-mail: martynov@bitp.kiev.ua},
 E. PREDAZZI$^{b,}$\footnote{\it E-mail: predazzi@to.infn.it}
 and A. PROKUDIN$^{b,c,}$\footnote{\it E-mail: prokudin@to.infn.it}}
\vskip 0.5cm
{\small\it
(a) Bogolyubov Institute for Theoretical Physics,\\ National Academy of
Sciences of Ukraine, \\ 03143 Kiev-143, Metrologicheskaja 14b, UKRAINE}
\vskip 0.2cm

{\small\it
\vskip 0.2cm
(b)  Dipartimento di Fisica Teorica,\\
Universit\`a Degli Studi Di Torino, \\
Via Pietro Giuria 1,
10125 Torino, \\
ITALY\\
and\\
Sezione INFN di Torino,\\
 ITALY\\}
\vskip 0.2cm
{\small\it
(c) Institute For High Energy Physics,\\
142281 Protvino,  RUSSIA}
\vskip 0.5cm
\parbox[t]{12.cm}{\footnotesize A model based on a dipole Pomeron framework
for vector meson exclusive photoproduction by real and virtual photons
shows very good agreement with the experimental data. The model does not
violate unitarity constrains and describes in a universal manner all the
available data for $\rho$, $\omega$, $\varphi$, $J/\psi$ and $\Upsilon$
vector meson photoproduction in the region of energies $1.7\le W \le
250\;GeV$ and photon virtualities $0 \le Q^2 \le 35\; GeV^2$. }
\end{center}

\section{Introduction}

Exclusive vector meson production by real and virtual photons is deemed to
provide important information on the transition region from the ``soft''
dynamics at low virtualities of the photon $Q^2$ to the ``hard''
perturbative regime at high $Q^2$. An evidence of such a transition was
the rapid growth of $J/\psi$ meson production discovered at HERA which
raised the question of the applicability and validity of the Regge pole
model in such a region.

Similarly the question of how to reconcile within QCD the transition from
the perturbative regime (hard diffraction) and the one of hadron-hadron
scattering cross-sections (soft diffraction) remains still quite open.

The effect discovered at HERA caused a vivid response by theorists.
Vector meson photoproduction is described in phenomenological and QCD based
approaches.

Based on the idea of soft Pomeron exchange, a model of Donnachie and
Landshoff~\cite{ref:DL} is capable of describing $\rho_0$ and $\varphi$
meson photoproduction using vector meson dominance with a scale factor
$0.84$. At the same time it fails to describe $J/\psi$ meson
photoproduction cross section since the modest rise of soft Pomeron
exchange with the intercept $\Delta_{\Pom}\simeq 0.08$ could not account
for the steep behaviour of the $J/\psi$ cross section. To cope with such a
steep behaviour they proposed a model of two Pomerons~\cite{ref:DL1} in
which an additional hard Pomeron is present with a high intercept
$\Delta_{\Pom_0}\simeq 0.4$ and the introduction of such an exchange helps
to describe the data on $J/\psi$ photoproduction and charm component
$F_2^c(x,Q^2)$.

The model of Haakman, Kaidalov and Koch \cite{ref:Kaidalov}, taking into
account both (`eikonal') multiple-Pomeron exchanges and a more complete
set of diagrams including the interactions among the exchanged Pomerons,
is capable of describing not only the photoproduction of vector mesons but
also the $Q^2$ dependent $\rho_0$ meson photoproduction by virtual
photons. The model is based on a Pomeron with intercept
$\Delta_{\Pom}\simeq 0.2$ which is softened by multi Pomeron exchanges in
hadron-hadron interactions and the effective intercept grows with $Q^2$
leading to an increase of the growth of vector meson cross sections with
increasing $Q^2$. The authors relate the vector meson differential cross
section to the proton structure function $F_2^p(x,Q^2)$ and obtain a good
description of the $\rho_0$ meson photoproduction by virtual photons using
their model for $F_2^p(x,Q^2)$ \cite{ref:Kaidalov1}.

Using a method of off shell extension of Regge eikonal model, Petrov and
Prokudin \cite{ref:Petrov} described $\rho_0$ and $J/\psi$ vector meson
production by real and virtual photons using the universal Pomeron with
intercept $\Delta_{\Pom_0}\simeq 0.1$ obtained from the description of $pp$
and $\bar p p$ total, elastic and differential cross sections. The model
uses a non $Q^2$ dependent Pomeron trajectory, takes into account
unitarity corrections and does not violate unitarity for hadron-hadron
processes and vector meson production.

A model based on dipole Pomeron by Jenkovszky, Martynov and Paccanoni
 \cite{ref:dipole_vector} describes the vector meson photoproduction cross
section for $\rho_0$, $\omega$, $\varphi$ and $J/\psi$ vector meson using
the Pomeron with intercept equal to 1, thus not violating unitarity
restrictions.

The Dipole Pomeron was used in Ref. \cite{ref:Jenk} to
describe heavy meson photoproduction. The model \cite{ref:Jenk} does
not violate unitarity and shows good agreement with the data.

At the same time vector meson photoproduction at large $Q^2$ is seen as a
test of perturbative QCD and of exchanges of the so called BFKL Pomeron.

Brodsky, Frankfurt, Gunion, Mueller and Strikman~\cite{ref:Brodsky} using
pQCD calculations show that with increasing $Q^2$, the vector meson
photoproduction cross section starts rising more steeply as compared with
photoproduction by real photons. They argue that the cross section is
dominated (at large $Q^2$) by two gluon exchange through its longitudinal
component
\be
\sigma\sim\sigma_L\sim \frac{1}{Q^6}[xg(x,Q^2)]^2\; ,
\ee
(where $xg(x,Q^2)$ is the gluon distribution) with respect to which the
transverse component is suppressed by a factor of $Q^2$ \be \sigma_T\sim
\frac{1}{Q^8}[xg(x,Q^2)]^2\; . \ee Martin, Ryskin and Teubner
\cite{ref:Martin} show that the diffractive electroproduction of $\rho$
meson at high $Q^2$ can be described by pQCD and argue that the conflict
between the QCD of prediction for the ratio
\be
\frac{\sigma_L}{\sigma_T}\sim\frac{Q^2}{2 m_\rho^2}
\ee
and the existing data \cite{ref:Rdata} may be resolved introducing the
relation
\be
\frac{\sigma_L}{\sigma_T}=\frac{Q^2}{M^2}\Big(\frac{\gamma (Q^2)} {\gamma
(Q^2)+1}\Big)^2,
\ee
where $M$ is the invariant mass of the $q\bar q$ pair
and $\gamma$ is the effective anomalous dimension of the gluon.

Cudell and Royen \cite{ref:Cudell} using the lowest-order QCD calculation
show that it reproduces correctly the ratios of cross sections for $\rho$,
$\varphi$ and $J/\psi$. At the same time, the data for
$\frac{\sigma_L}{\sigma_T}$ are not reproduced as in most of the models
and they propose a new approach \cite{ref:Cudell1} where the asymptotic
form of the transverse cross section gives $\sigma_T\sim\frac{1}{Q^6}$
leading to a good description of the ratio $\frac{\sigma_L}{\sigma_T}$.

The energy dependence of the vector meson photoproduction cross section
was investigated in a paper by Donnachie, Gravelis and Shaw
\cite{ref:Donnachie} who use pQCD calculations, the model of Cudell and
Royen \cite{ref:Cudell1}, soft Pomeron contribution modelling
non-perturbative two-gluon exchange and several models for the vector
meson wave function and achieve good global description of the vector
meson data.

Nemchik, Nikolaev, Predazzi and Zakharov \cite{ref:Nemchik} use color
dipole phenomenology and predict for the dipole cross section
\be
\frac{\sigma(\gamma^*\rightarrow\omega)}
{\sigma(\gamma^*\rightarrow\rho_0)}=\frac{1}{9}
\ee
independent of energy and $Q^2$. The elastic dipole cross
section is in accord with the QCD prediction
\be
\sigma(\gamma^*\rightarrow V) = \sigma_T(\gamma^*\rightarrow V)+ \epsilon
\sigma_L(\gamma^*\rightarrow V)\sim \frac{1}{(Q^2+m_V^2)^4}\Big(
1+\epsilon R_{LT}\frac{Q^2}{m_V^2}\Big).
\ee
Nemchik, Nikolaev and
Zakharov \cite{ref:Nemchik1} observe that elastic production of vector
mesons is of great potentiality for probing the BFKL Pomeron
\cite{ref:BFKL}.

Vanderhaeghen, Guichon and Guidal \cite{ref:Guidal}
estimate the leading order amplitudes for exclusive
meson electroproduction at large $Q^2$ in terms of
skewed quark distributions and obtain a good description of
$\sigma_L$ for $\rho_0$ meson exclusive production.

The $|t|$ dependence of $J/\psi$ production is discussed in
\cite{ref:Levin} using DGLAP equation with one
and two Pomeron exchanges being taken into account.

High $t$ vector meson production, finally, is analyzed in a paper by Laget
\cite{ref:Laget}. Good description of elastic cross section for $\rho$,
$\omega$ and $\varphi$ meson is obtained using the soft Pomeron approach,
but the $J/\psi$ cross section is not well reproduced.

The problem of reconciling
``soft'' and ``hard'' regimes is still present despite the substantial
progresses of the theory. It is not clear if the hard Pomeron is ultimately
needed to describe the data or not (see for example
\cite{ref:Desgrolard}). We attempt to
describe the data in the framework of ``soft'' physics, so that not
we do not violate unitarity. In addition, we describe
hadron-hadron processes thus answering in  the positive
the question: is it possible to use the universal soft Pomeron not
only for hadron-hadron processes but also for photoproduction of light and
heavy vector mesons?

\section{General formalism}
We utilize the following picture of the interaction: a photon fluctuates
into a quark-antiquark pair and as the lifetime of such a fluctuation is
quite long (by the uncertainty principle it grows with the beam energy
$\nu$ as $2\nu/(Q^2+M_V^2)$ \cite{ref:ioffe}), the proton interacts via
Pomeron or secondary Reggeon exchange with this quark-antiquark pair.
After the interaction this pair forms a vector meson \cite{ref:Brodsky}.
We may conclude that such an interaction must be very close to that among
hadrons and following the principle of Regge pole theory that the Pomeron
is universal in all hadron-hadron interactions, we use the same Pomeron
which is used to describe hadron-hadron interaction. Thus if Pomeron
exchange is present in some interaction, then it has the same properties
(is the form of singularity, position of such a singularity in the
$J$-plane, trajectory etc.) as in hadron-hadron interaction. This is true
at least for on shell particles. Concerning the $Q^2=0$ photon, we
consider it as a hadron (according to the experimental data). Then for
$Q^2\neq 0$ we assume that no new singularity appears
\cite{ref:Evolution}. More precisely, even if we assume a new singularity
at $Q^2\neq 0$, its contribution must be equal to zero for $Q^2= 0$.
Indeed the analysis of the data \cite{ref:Desgrolard} shows that there is
no need for such a new contribution.

Our choice of the Pomeron contribution is the so called dipole Pomeron
which gives a very good description of all hadron-hadron total cross
sections~\cite{ref:dipole_hadron} and was used already to describe
photoproduction of vector mesons~\cite{ref:dipole_vector}. However the
dipole Pomeron was applied in this paper only at $Q^{2}=0$ in a
qualitative rather than in a detailed description
 of the data. In addition, a large wealth of new data
have appeared since \cite{ref:dipole_vector}.

\begin{figure}[h]
\begin{center}
\includegraphics*[scale=0.8]{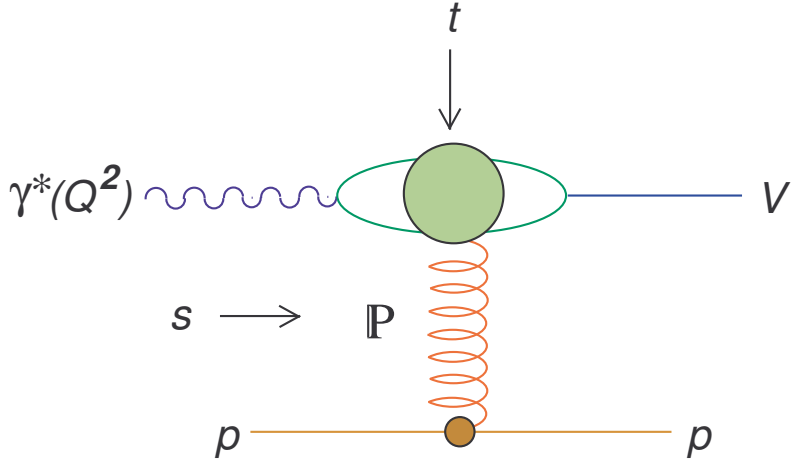}
\caption{Photoproduction of a vector meson.}
\label{Figure 1}
\end{center}
\end{figure}

The basic diagram is depicted in Figure \ref{Figure 1}; $s$ and $t$ are
the usual Mandelstam variables, $Q^2=-q^2$ is the virtuality of the
photon.

We introduce the ``scaling'' variable $\tilde Q^2$
$$ \tilde Q^2= Q^2+M_V^2\;,$$
where $M_V^2$ is the mass of the vector mesons. One assumes that a large
meson mass plays the same r\^ole as the photon virtuality setting a
``hard'' scale for the reaction.

In general, the dipole Pomeron model generalized for virtual
external particles can be written as \cite{ref:JMP}
\be
A(W^2,t;\tilde Q^2) = \Pom (W^2,t;\tilde Q^2) + f(W^2,t;\tilde Q^2)
+ ...\;,
\ee
where  $W^2=(p+q)^2=m_p^2+2 m_p \nu - Q^2$.
$\Pom$ is the Pomeron contribution, with $s=W^{2}$
\be
\Pom (W^2,t;\tilde Q^2) =
ig_{0}(t;\tilde Q^2) \Bigl(\frac{-is}{s_{0}(\tilde
Q^2)}\Bigr)^{\alpha_{\Pom}(t)-1} + ig_{1}(t;\tilde
Q^2)ln(\frac{-is}{s_{1}(\tilde Q^2)}) \Bigl(\frac{-is}{s_{1}(\tilde
Q^2)}\Bigr)^{\alpha_{\Pom}(t)-1}
\ee
 and
\be
g_{i}(t;\tilde Q^2) =
g_{i}(\tilde Q^2)\exp (b_{i}(\tilde Q^2)t).
\ee
A similar expression applies to the contribution of
the $f$-Reggeon
\be
f(W^2,t;\tilde Q^2) =
ig_{f}(t;\tilde Q^2) \Bigl(\frac{-is}{s_{f}(\tilde
Q^2)}\Bigr)^{\alpha_{f}(t)-1}.
\ee
It is important to stress that in this model the intercept of the
Pomeron trajectory is equal to 1
\be
\alpha_{\Pom}(0) = 1.
\ee
Thus the model does
not violate the Froissart-Martin bound ~\cite{ref:MartinF}.

The parameters in $\alpha_{\Pom}(t)$ and $\alpha_{f}(t)$ are universal and
independent of the reaction; $g_{i},b_{i},s_{i}$ are functions of the
variable $\tilde Q^2$ and the same for all reactions. Universality is a
good approximation, which turns out to work very well in the present case,
though in the framework of Regge theory strict universality does not hold
and, in principle, most parameters could depend on the type of reaction as
well. It may happen that new detailed set of data will need such a
dependence; this of course would not automatically imply that the model is
not able to describe the data or that the framework is a false one.

In hadronic phenomenology $g_{i}$ and
$b_{i}$ are constants. Here, they may, in general, be
$Q^{2}$-dependent functions. The slope
$B(s,t;Q^{2})=2b_{i}(Q^{2})+2\alpha^{'}_{\Pom}ln(s/s_{i}(Q^2))$
contains a universal energy-dependent term,
while the parameter $b_{i}(Q^{2})$ is responsible for the quark content.
In what follows we find that the experimental data do not demand
any $Q^2$ dependence in $b_{i}(Q^{2})$; as a consequence
we use $Q^2$ independent
parameters $b_{i}$, where $i=\Pom, f$.

Let us stress here that the only variable that differentiates among the
various vector meson elastic cross sections in this framework is the {\it
mass} of the vector mesons in the case of photoproduction ($Q^2=0$) or
$\tilde Q^2$, i.e. the sum of the {\it mass} and {\it virtuality} of the
photon in the case of DIS scattering. Thus, we propose a universal
description for all vector mesons. We will see that such a description is
very good, for example, we {\it predict} the correct cross sections for
heavy vector meson $\Upsilon$ in the case of photoproduction, and the
correct $Q^2$ and $W^2$ dependence for $\omega$, $\varphi$ and $J/\psi$
vector mesons in the case of DIS when we fit only the $\rho$ meson cross
sections. The differential cross sections in all cases are described
without any additional fitting. Such set of predictions give credit to the
present model, given that most approaches have been able to make similar
predictions in a correct way.

\section{The Model}
\subsection{Photoproduction of vector mesons by real photons}

For $\rho$, $\varphi$ and $J/\psi$ meson photoproduction we write the
scattering amplitude as the sum of a Pomeron and $f$ contribution.
According to Okubo-Zweig rule, the $f$ meson contribution ought to be
suppressed in the production of $\varphi$ and $J/\psi$ mesons, though,
taking into account the present crudeness of the state of the art, we
added the $f$ meson contribution even in the $J/\psi$ meson case. The
result of the fit proved that the $f$ contribution is indeed negligible
for $J/\psi$ meson production (see the end of this Section for details)
whereas it is not irrelevant for $\varphi$ meson production. We believe,
in fact, that due to $\omega - \phi$ mixing $f$ contribution is present in
the latter; indeed in the $\varphi$ decay mode, more than $15$\% is due to
non strange particles and the $\bar K K$ decay mode is present for $f$
meson decay. Thus, we conclude that the $f$ meson may contribute to
$\varphi$ meson production.

For $\omega$ meson photoproduction, we include also $\pi$ meson exchange
(see also the discussion in \cite{ref:DL}), which is needed to describe
the low energy data given that we try to describe the data for all
energies $W$, starting from its threshold. Initially we did not expect
model to work in the threshold area but it turned out to be capable of
describing well the reactions even around threshold.

By writing the amplitude as $A=A_{\Pom}+A_{\Reg}$, where $\Reg$ stands for
the secondary Reggeon contribution, we obtain for the integrated elastic
cross section $\sigma_{el}$

\be
\sigma(W^2, M_V^2)^{\gamma p\rightarrow
Vp}_{el} = 4\pi\int\limits_{-\infty}^{0}dt|A^{\gamma p\rightarrow
Vp}(W^2,t;M_V^{2})|^{2}\; .
\label{eq:sigma}
\ee

We propose the following parametrizations for Pomeron and Reggeon
couplings:
\bea\label{eq:couplings}
g_0(t; M_V^2)=\frac{g_0 M_V^2}{(W_0^2+M_V^2)^2}exp(b_\Pom^2 t)\; ; \\
\nonumber
g_1(t; M_V^2)=\frac{g_1 M_V^2}{(W_0^2+M_V^2)^2}exp(b_\Pom^2 t)\; ; \\
\nonumber
g_\Reg(t; M_V^2)=\frac{g_\Reg M_p^2}{(W_0^2+M_V^2)^2}exp(b_\Reg^2 t)\; , \\
\nonumber
\eea
where $g_0, \; g_1\;,$ $W_0^2\; (GeV^2)\;$, $b^2_\Pom \;
(GeV^{-2})\;$ are adjustable parameters; $M_p^2$ is the proton mass.
$\Reg=f$ for $\rho$, $\varphi$ and $J/\psi$, and $\Reg=f,\pi$ for $\omega$.
$g_f$, $g_\pi$, $b^2_\Reg\; (GeV^{-2})\;$ are also adjustable parameters.
We use the same slope $b^2_\Reg$ for $f$ and $\pi$ Reggeon exchanges.

The parametrizations of the Pomeron contribution is the
following
\bea\label{eq:aPomeron}
A_{\Pom}(W^2,t;M_V^2) =ig_0(t; M_V^2)\Big(-i\frac{W^2-M_p^2}{W_0^2+M_V^2}
\Big)^{\alpha_\Pom (t)-1}+ \\
\nonumber
ig_1(t; M_V^2)\ln \Big(-i\frac{W^2-M_p^2}{W_0^2+M_V^2}\Big)
\Big(-i\frac{W^2-M_p^2}{W_0^2+M_V^2}\Big)^{\alpha_\Pom (t)-1}\; ,
\eea
 where we use a linear Pomeron trajectory
\be
\alpha_\Pom (t)=1+\alpha'_\Pom (0)\: t\;
\ee
with $\alpha'_\Pom (0)=0.25\; (GeV^{-2})$.

The contribution of the secondary Reggeons to the amplitude is written as
\be\label{eq:aReggeon}
A_{\Reg}(W^2,t;M_V^2) =ig_\Reg(t; M_V^2)\Big(-i\frac{W^2-M_p^2}
{W_0^2+M_V^2}
\Big)^{\alpha_\Reg (t)-1}\; .
\ee

In order to take into account the threshold behaviour \footnote{ We believe
that  without such a factor it is not possible to 
obtain the correct parameters for the secondary Reggeons (and 
correspondingly 
those of the Pomeron), and thus a correct
description of the data for $J/\psi$ meson production
at high energies $W$ (let alone the data near threshold).} we
multiply the amplitude $A$ by a threshold factor \be \label{eq:threshold}
\Big(1-\frac{(M_p+M_V)^2}{W^2}\Big)^{\sqrt{M_V^2/M_0^2}} \ee where
$M_0^2\;(GeV^2)$ is an adjustable parameter and $M_p+M_V$ is the reaction
threshold. The threshold behaviour of the amplitude is also discussed in
\cite{ref:Brodsky2} where the exponent ( calculated in terms of the number
of quark spectators $n_s$) is $2n_s$, where $n_s=0,1,\; {\rm or}\; 2$. The
empirical value we find in our model, $1.3$ for the $J/\psi$ threshold
behaviour (see Table \ref{Table 1.}), is well compatible with the result of
\cite{ref:Brodsky2}.

We take into account the isostructure of vector mesons using the following
relation between $\omega$ and $\rho_0$ cross-sections ($\omega$ is an
isoscalar while $\rho_0$ is an isovector) which is predicted to hold in
 \cite{ref:Nemchik}
\be
\frac{\sigma_{\omega}}{\sigma_{\rho}} =
\frac{1}{9}\; .
\ee
Accordingly, given that the masses of $\omega$ and
$\rho_0$ are very close one to the other, we multiply the cross-section in
Eq. \ref{eq:sigma} by a factor of $9$ for the case of $\rho_0$ meson
exclusive production.

In the fit we use all available data starting from the threshold for each
meson and we do not discriminate among sets of data having different
normalizations. It is, however, quite evident that, especially at low
energies, different experiments have different normalizations. This
implies that the $\chi^2/{\rm d.o.f.}$ will not be very good while the
overall agreement is quite satisfactory.

The whole  set of data is composed of $272$ experimental points
\footnote{The data are available at \\
REACTION DATA Database {\it http://durpdg.dur.ac.uk/hepdata/reac.html} \\
CROSS SECTIONS PPDS database {\it http://wwwppds.ihep.su:8001/c1-5A.html}}
 and having
a grand total of $8$ parameters, we find $\chi^2/{\rm d.o.f}=1.32$. The
main contribution to the $\chi^2$ comes from the low energy region ( $W\le
4\;GeV$); had we started fitting from $W_{min}=4\;GeV$, the resulting
$\chi^2/{\rm d.o.f}=0.96$ would be much better and more appropriate for a
high energy model.

The parameters are given in Table \ref{Table 1.}. The errors of the
parameters are those obtained by MINUIT.
\begin{table}[H]
\begin{center}
\begin{tabular}{|l|l|r|r|}
\hline
N & Parameter & Value & Error \\
\hline
1 & $g_1$ & 0.18487E-01&   0.67523E-03 \\
2 & $g_0$ &-0.45223E-01&   0.24077E-02 \\
3 & $g_f$ &  0.19866   &    0.43331E-02 \\
4 & $g_\pi $& 0.38904  &     0.10219E-01 \\
5 & $b^2_\Pom\; (GeV^{-2})$ &0.51532 &      0.10462 \\
6 & $b^2_f\; (GeV^{-2})$ &   0.76365 &      0.62196E-01 \\
7 & $W_0^2\; (GeV^{2})$&   0.90696 &      0.10059E-01\\
8 & $M_0^2\; (GeV^{2})$&   5.6044  &     0.45157  \\
\hline
\hline
& Trajectory & $\alpha (0)$ (FIXED)& $\alpha' (0)\; GeV^{-2}$ (FIXED) \\
\hline
1&Pomeron & 1 & 0.25  \\
2&$f$ Reggeon & 0.8 & 0.85 \\
3&$\pi$ Reggeon & 0.0  & 0.85 \\
\hline
\hline
& Meson & \# of points & $\chi^2$ per point \\
\hline
1 & $\rho_0(770)$ & 129 & 1.47 \\
2 & $\omega(782)$& 56  & 1.51 \\
3 & $\phi(1020)$ & 38  & 0.82 \\
4 & $J/\Psi(3096)$&49  & 0.84 \\
\hline
\hline
& All mesons & \# of points & $\chi^2/{\rm d.o.f.}$  \\
\hline
&$\rho_0,\omega,\phi,J/\Psi $ & 272 & 1.32 \\
\hline
\end{tabular}
\end{center}
\vskip -0.5cm
\caption{ Parameters obtained by fitting $\rho_0,\omega,\phi$ and $J/\psi$
photoproduction data \label{Table 1.}}
\end{table}

\begin{figure}[H]
\centering
{\vspace*{ -1cm} \epsfxsize=130mm \epsffile{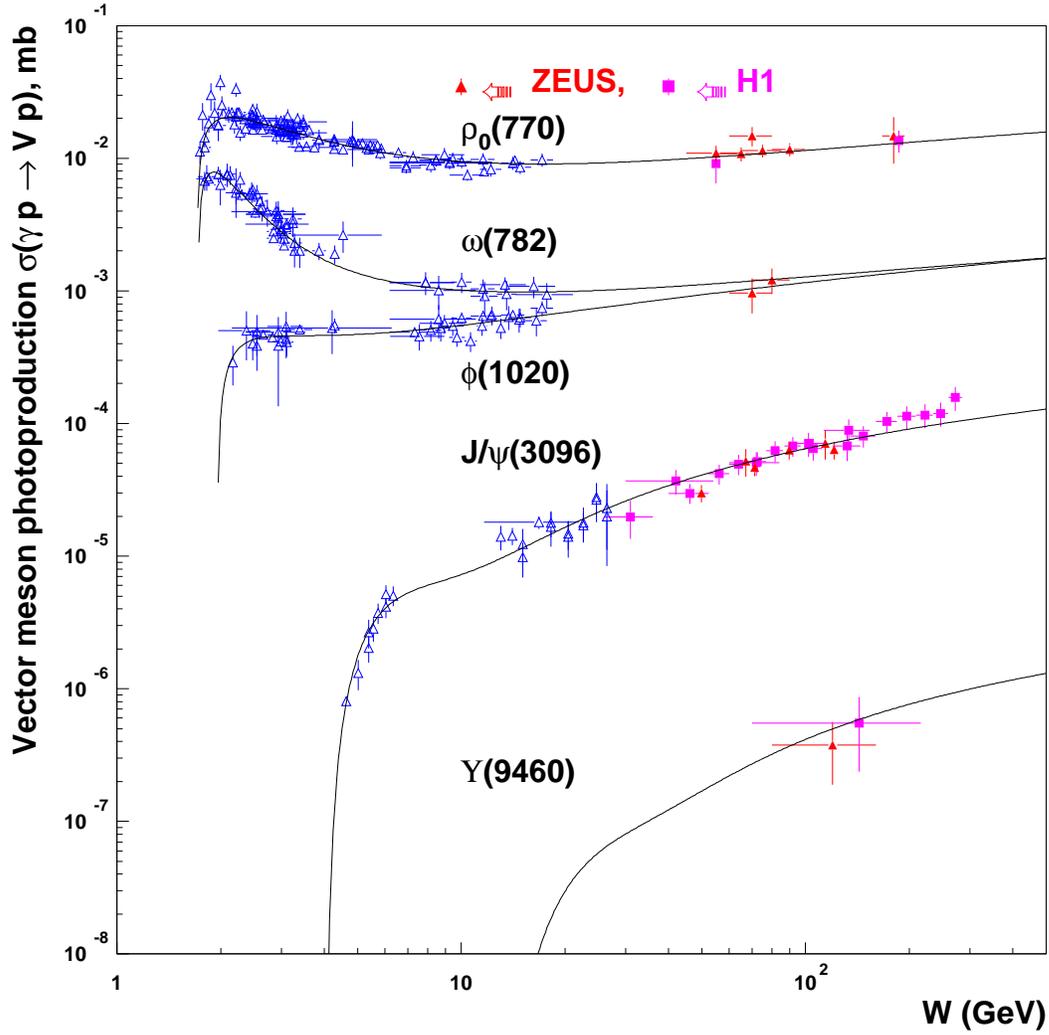}}
\vskip-2.cm
\caption{Elastic cross-sections of vector mesons photoproduction. The
curve for $\Upsilon (9460)$ production is a prediction
of the model.
\label{fig:mesons}}
\end{figure}

The results are presented in Fig. \ref{fig:mesons}, which shows also the
 prediction of the model for $\Upsilon (9460)$ photoproduction.

As a temporary comment, we stress that the model describes the data on
vector meson exclusive photoproduction without the need of Pomeron
contribution with intercept higher than 1.

Moreover, the rapid rise of the $J/\psi$ cross section at low energies is
described as a transition phenomenon, a delay of the onset of the real
asymptotics.

Most remarkable, the data on $\Upsilon (9460)$ production is very
well {\it predicted} in this framework, as seen in Fig. \ref{fig:mesons}.

Let us address once more the issue of $J/\psi$ meson photoproduction. In
the framework of the model, the rapid growth of the cross section is a
transition effect of the onset of the asymptotic behavior.

\begin{figure}[H]
\parbox[c]{8.6cm}{\epsfxsize=76mm
\epsffile{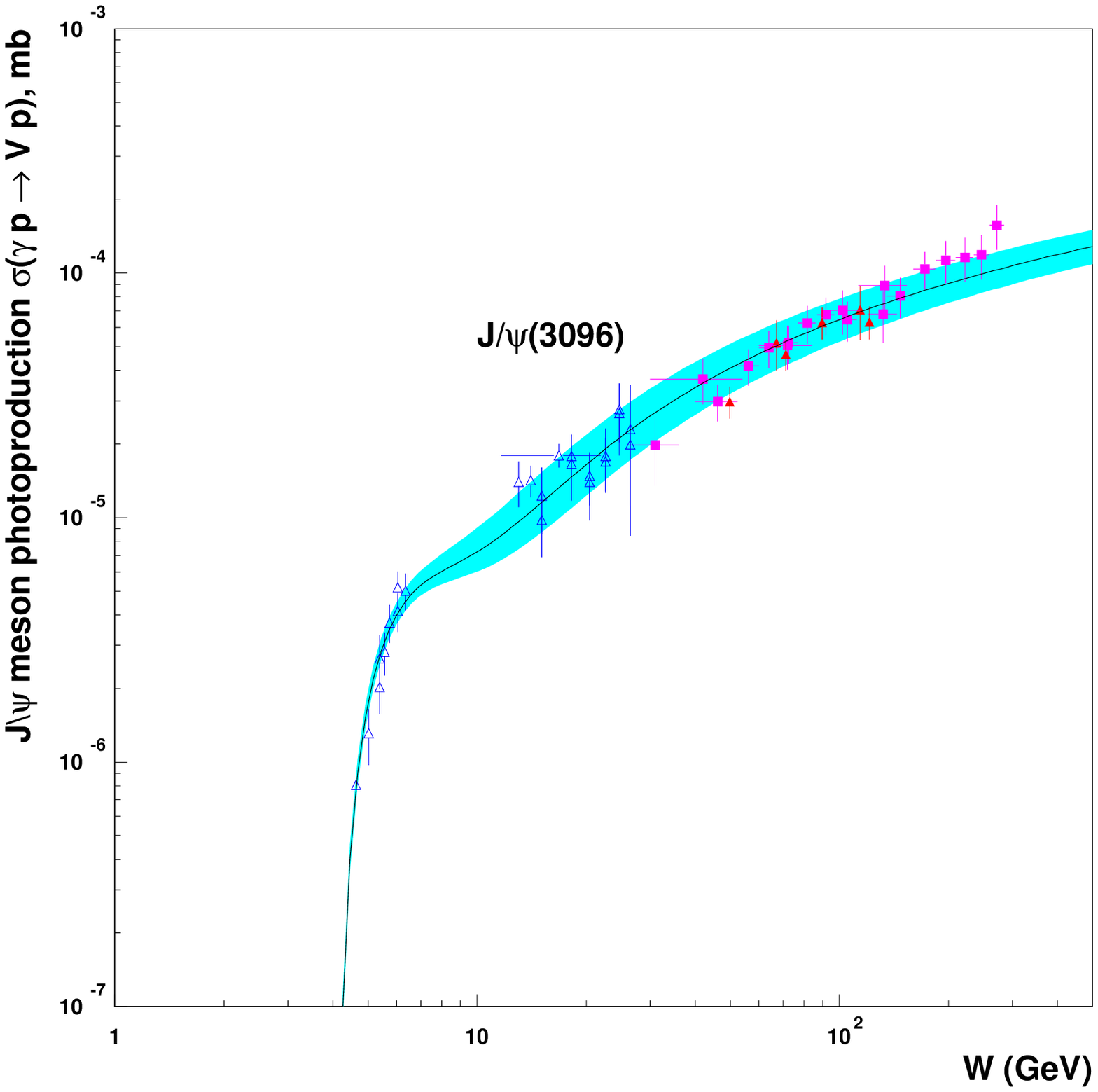}} \hfill~\parbox[c]{7.6cm}{\epsfxsize=76mm
\epsffile{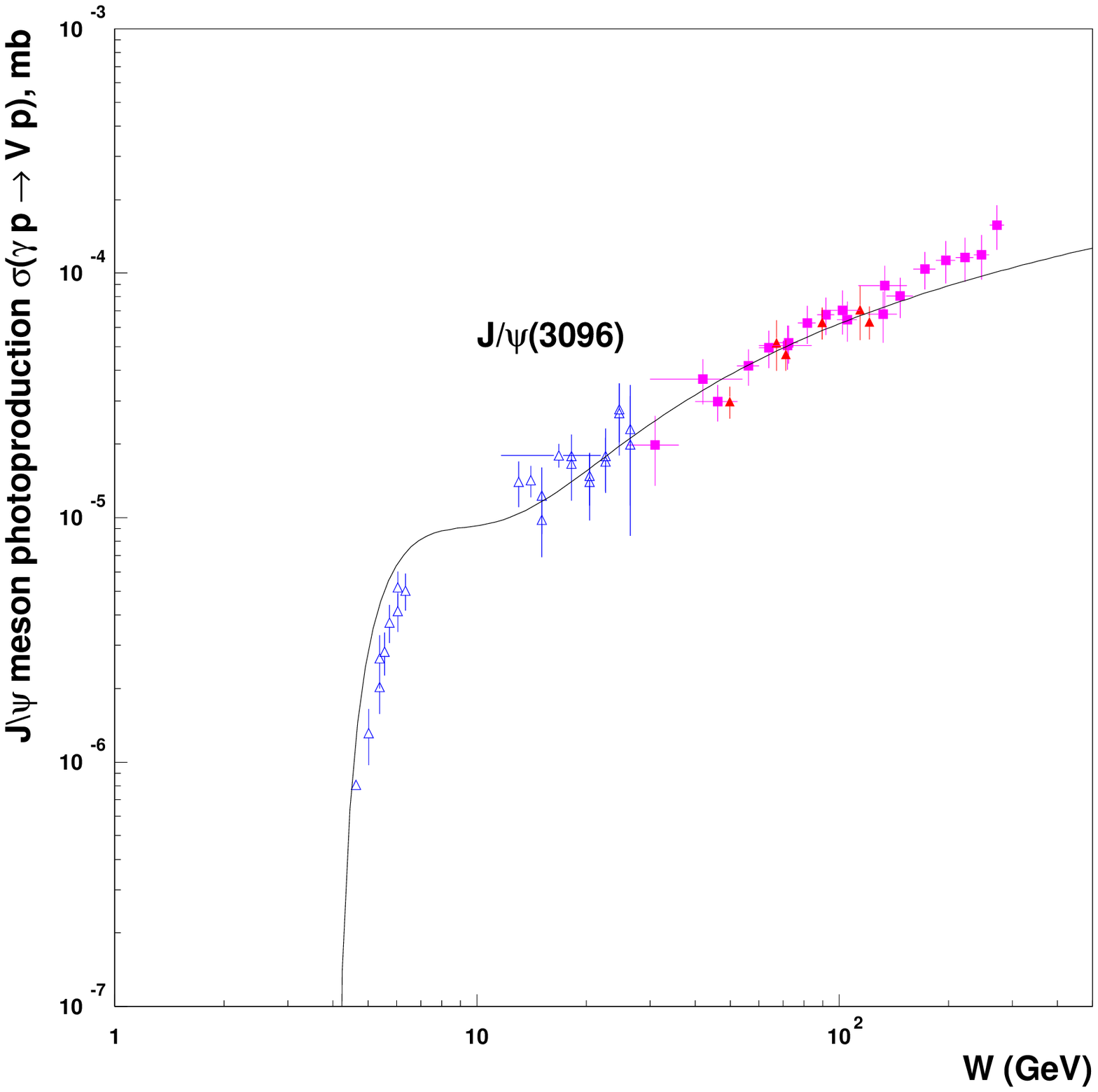}}

\vspace*{-1.7cm}
\parbox[t]{7.7cm}{\caption{Elastic cross section of exclusive $J/\psi$
meson photoproduction. Shaded area represents the uncertainty corridor
due to the errors of parameters .\label{fig:jpsi0}}}
\hfill~\parbox[t]{7.7cm}{\caption{Elastic cross section of exclusive
$J/\psi$ meson photoproduction without $f$ Reggeon contribution.
\label{fig:jpsi0f}}}
\end{figure}

In Fig. \ref{fig:jpsi0} one can see the cross section and the available
corridor due to the uncertainty in the parameters (see Table \ref{Table
1.}). We suppose that with the growth of $W$, the energy dependence of
elastic cross section will change to a mild $\ln (s)$ behavior thus
growing no faster than hadron-hadron elastic cross-sections. Another
interesting question is the suppression of secondary Reggeons. As the
quark content of $J/\psi$ ($\bar c c$) suggests that $f$ Reggeon exchange
(composed of light quarks) must be suppressed, we show in Fig.
\ref{fig:jpsi0f} corresponding cross section without $f$ Reggeon
contribution. We may conclude that in the model we have this suppression,
as the contribution of the $f$ Reggeon is just some percent in the region
of low $W$ and negligible for $W\ge 20\; GeV$.

\subsection{Photoproduction of vector mesons by virtual photons}

In the case of nonzero virtuality of the photon, we have a new variable in
play $Q^2=-q^2$. At the same time, we have a nonzero contribution of
$\sigma_L$ for the cross section. According to \cite{ref:Brodsky}, QCD
predicts the following dependence for $\sigma_T$, $\sigma_L$ and their
ratio as $Q^2$ goes to infinity: \bea \nonumber
\sigma_T \sim \frac{1}{Q^8}(x_\Pom g(x_\Pom, \tilde Q^2))^2 ; \\
\label{eq:qcd}
\sigma_L \sim \frac{1}{Q^6}(x_\Pom g(x_\Pom, \tilde Q^2))^2 ; \\
\nonumber
R\equiv \sigma_L/\sigma_T \sim Q^2/M_V^2; \\
\nonumber
\sigma = (\sigma_T + \sigma_L) |_{Q^2 \rightarrow \infty} \sim \sigma_L.
\nonumber
\eea
Here $x_\Pom g(x_\Pom, \tilde Q^2)$ is the gluon distribution function,
$\tilde Q^2\equiv \frac{Q^2+M_V^2}{4}$, $x_\Pom=\frac{Q^2+M_V^2}
{W^2+M_V^2}$.

Enforcing these predictions, we use the following (most economical)
parametrization for $R$ (which cannot be deduced from the Regge theory)
\be
R(Q^2, M_V^2) = c\frac{Q^2+M_V^2}{Q_r^2+Q^2+M_V^2}\frac{Q^2}{M_V^2}
\label{eq:ratio}
\ee
where $c$ and $Q_r^2\; (GeV^2)$ are adjustable
parameters. Thus, the asymptotic behaviour of $R(Q^2, M_V^2)$ reproduces
that of Eq. \ref{eq:qcd}
\be
R(Q^2, M_V^2)\big|_{Q^2\rightarrow \infty }
\sim \frac{Q^2}{M_V^2}\; .
\ee

Accordingly, in the case $Q^2\neq 0$ we use the following parametrizations
for Pomeron and Reggeons couplings (compare with Eq. \ref{eq:couplings}):
\bea\label{eq:couplingsq}
g_0(t; Q^2, M_V^2)=\frac{g_0 M_V^2}{(W_0^2+Q^2+M_V^2)^2}
\sqrt{\frac{Q_0^2}{Q_0^2+Q^2}}exp(b_\Pom^2 t)\; ; \\
\nonumber
g_1(t; Q^2, M_V^2)=\frac{g_1 M_V^2}{(W_0^2+Q^2+M_V^2)^2}
exp(b_\Pom^2 t)\; ; \\
\nonumber
g_\Reg(t; Q^2, M_V^2)=\frac{g_\Reg M_p^2}{(W_0^2+Q^2+M_V^2)^2}
\sqrt{\frac{Q_\Reg^2}{Q_\Reg^2+Q^2}}exp(b_\Reg^2 t)\; , \\
\nonumber
\eea
where $Q_0^2\;(GeV^2)$ and $Q_\Reg^2\;(GeV^2)$  are  adjustable parameters.

In Eq. \ref{eq:couplingsq} we have introduced a new factor that
differentiates virtual from real photoproduction
\be
f(Q^2)=\sqrt{\frac{Q_{0,\Reg}^2}{Q_{0,\Reg}^2+Q^2}},\; f(0)=1\; .
\ee
Since in Eq. \ref{eq:couplings} we have complete control in the $Q^2$
dependence (recall that we use the sum $\tilde Q^2=Q^2+M_V^2$ as a
variable) up to any dimensionless function $f(Q^2)$, such that $f(0)=1$.

The parametrization of the Pomeron contribution is now the following
(compare with Eq. \ref{eq:aPomeron})
\bea\label{eq:aqPomeron}
A_{\Pom}(W^2,t;M_V^2,Q^2) =ig_0(t; Q^2, M_V^2)\Big(-i\frac{W^2+Q^2-M_p^2}
{W_0^2+Q^2+M_V^2}\Big)^{\alpha_\Pom (t)-1}+ \\
\nonumber ig_1(t; Q^2, M_V^2)\ln
\Big(-i\frac{W^2+Q^2-M_p^2}{W_0^2+Q^2+M_V^2}\Big)
\Big(-i\frac{W^2+Q^2-M_p^2}{W_0^2+Q^2+M_V^2}\Big)^{\alpha_\Pom (t)-1}\; ,
\eea
while the parametrization of the Reggeon contribution is (compare
with Eq. \ref{eq:aReggeon})
\be\label{eq:aqReggeon}
A_{\Reg}(W^2,t;M_V^2,Q^2) =ig_\Reg(t; Q^2,
M_V^2)\Big(-i\frac{W^2+Q^2-M_p^2}{W_0^2+Q^2+M_V^2}\Big)^{\alpha_\Reg
(t)-1}\; .
\ee

For simplicity, we multiply both components by the same threshold factor
already used for real photoproduction (Eq.~\ref{eq:threshold})
\be
\Big(1-\frac{(M_p+M_V)^2}{W^2}\Big)^{\sqrt{M_V^2/M_0^2}}\; .
\ee

The (integrated) elastic cross-section can now be written as
\be
\sigma^{\gamma^* p\rightarrow
Vp}_{el}(W^2;M_V^2,Q^2) = \sigma_{T}(W^2;M_V^2,Q^2)+
\sigma_{L}(W^2;M_V^2,Q^2)\; ,
\ee
where
\bea
\nonumber
\sigma_T(W^2;M_V^2,Q^2) = 4\pi\int\limits_{-\infty}^{0}dt
|A_\Pom(W^2,t;M_V^{2})+A_\Reg(W^2,t;M_V^{2})|^{2}\big|_{Q^2
\rightarrow \infty } \propto \frac{1}{Q^8}\; ; \\
\sigma_L(W^2;M_V^2,Q^2) = R(Q^2, M_V^2)\sigma_T(W^2;M_V^2,Q^2)
\big|_{Q^2\rightarrow \infty } \propto \frac{1}{Q^6}\; ;\\
\nonumber
\sigma = (\sigma_T + \sigma_L) \big|_{Q^2\rightarrow \infty } \sim
 \sigma_L\; .
\eea
Notice that we have directly enforced the asymptotical behaviour
expected from QCD (Eq.~\ref{eq:qcd}) for $\sigma_T$ and $\sigma_L$ through
the (empirical) choice of $R$ made earlier (Eq.~\ref{eq:ratio}).

We have, altogether 4 additional adjustable parameters as compared with
real photoproduction.

In order to obtain the values of the parameters for the case $Q^2\ne 0$,
we choose to fit just the data\footnote{The data are available at \\
REACTION DATA Database {\it http://durpdg.dur.ac.uk/hepdata/reac.html} \\
CROSS SECTIONS PPDS database {\it http://wwwppds.ihep.su:8001/c1-5A.html}}
on $\rho_0$ meson photoproduction in the region $0\le Q^2\le 35 \; GeV^2$;
the parameters for photoproduction by real photons are the same as in
Table \ref{Table 1.}. In order to avoid the low $W$ region where nucleon
resonances may spoil the picture of $\rho$ meson exclusive production, we
restrict the energy region to the domain $W\geq 4\; GeV$. The parameters
thus obtained are shown in Table \ref{Table 2.}.
\begin{table}[H]
\begin{center}
\begin{tabular}{|l|l|r|r|}
\hline
N & Parameter & Value & Error \\
\hline
1 & $c$ & 1.6569    &    0.27570 \\
2 & $Q_r\; (GeV^{2})$ & 7.2822   &     1.3013 \\
3 & $Q_0\; (GeV^{2})$ & 7.7921   &     4.9477 \\
4 & $Q_\Reg\; (GeV^{2})$ & 1.1209   &  0.69151E-01 \\
\hline
\hline
& Fit by a meson & \# of points & $\chi^2/{\rm d.o.f.}$  \\
\hline
&$\rho_0(770)$ & 200 & 1.33 \\
\hline
\hline
& Meson & \# of points & $\chi^2$ per point \\
\hline
1 & $\rho_0(770)$ & 200 & 1.30 \\
2 & $J/\Psi(3096)$& 75  & 0.83 {\bf (no fit!)}\\
\hline
\end{tabular}
\end{center}
\vskip -0.5cm
\caption{ Parameters obtained by fitting to $\rho_0$
photoproduction by virtual photons data \label{Table 2.}}
\end{table}

\begin{figure}[H]
\centering
{\vspace*{ -1cm}
\epsfxsize=150mm \epsffile{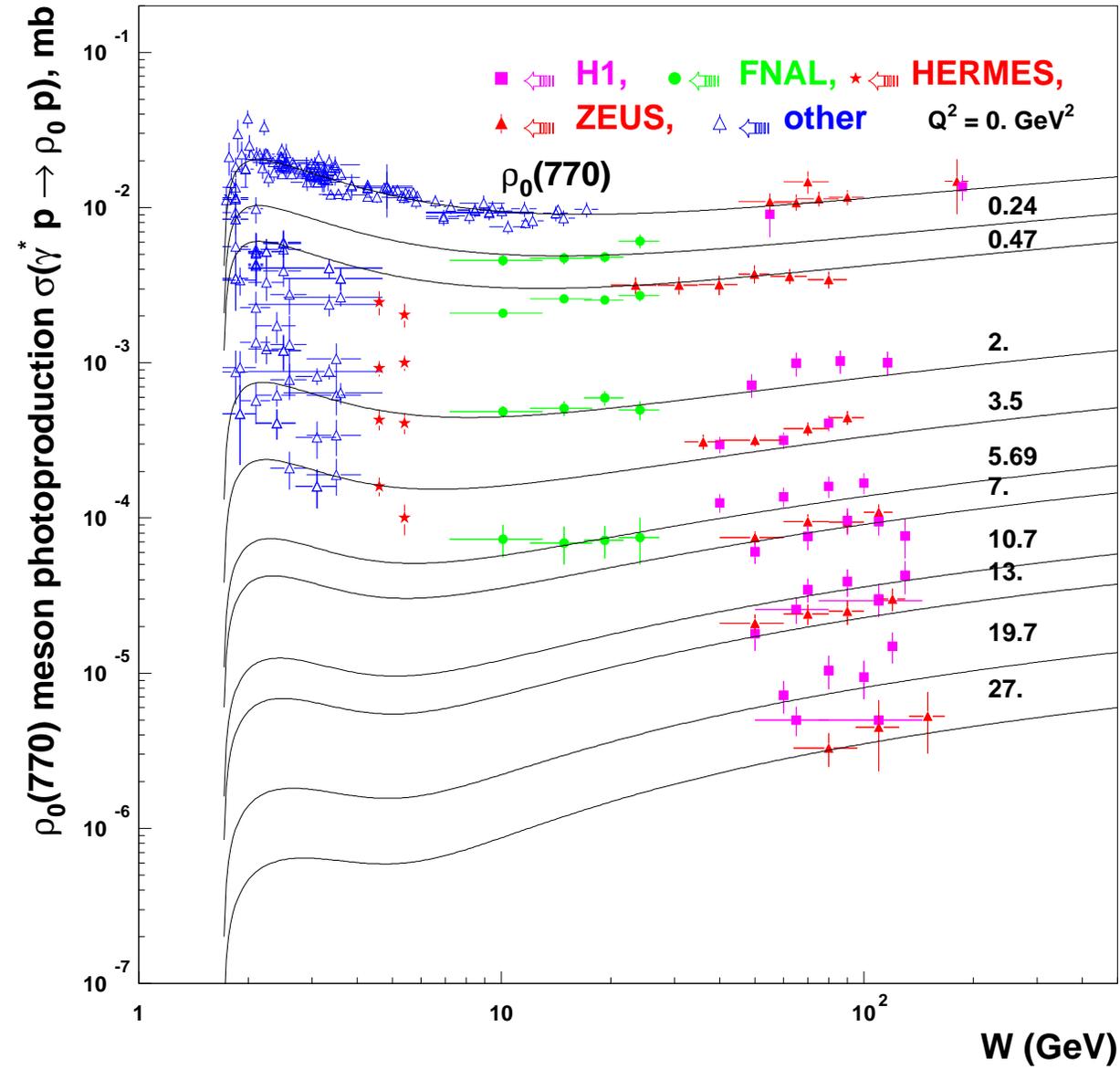}}
\vskip -3.cm
\caption{Elastic cross section of exclusive $\rho_0$ meson
photoproduction by virtual photons as a function of $W$ for different
values of $Q^2$. \label{fig:rho} }
\end{figure}

\begin{figure}[H]
\centering
{\vspace*{ -1cm}
\epsfxsize=150mm \epsffile{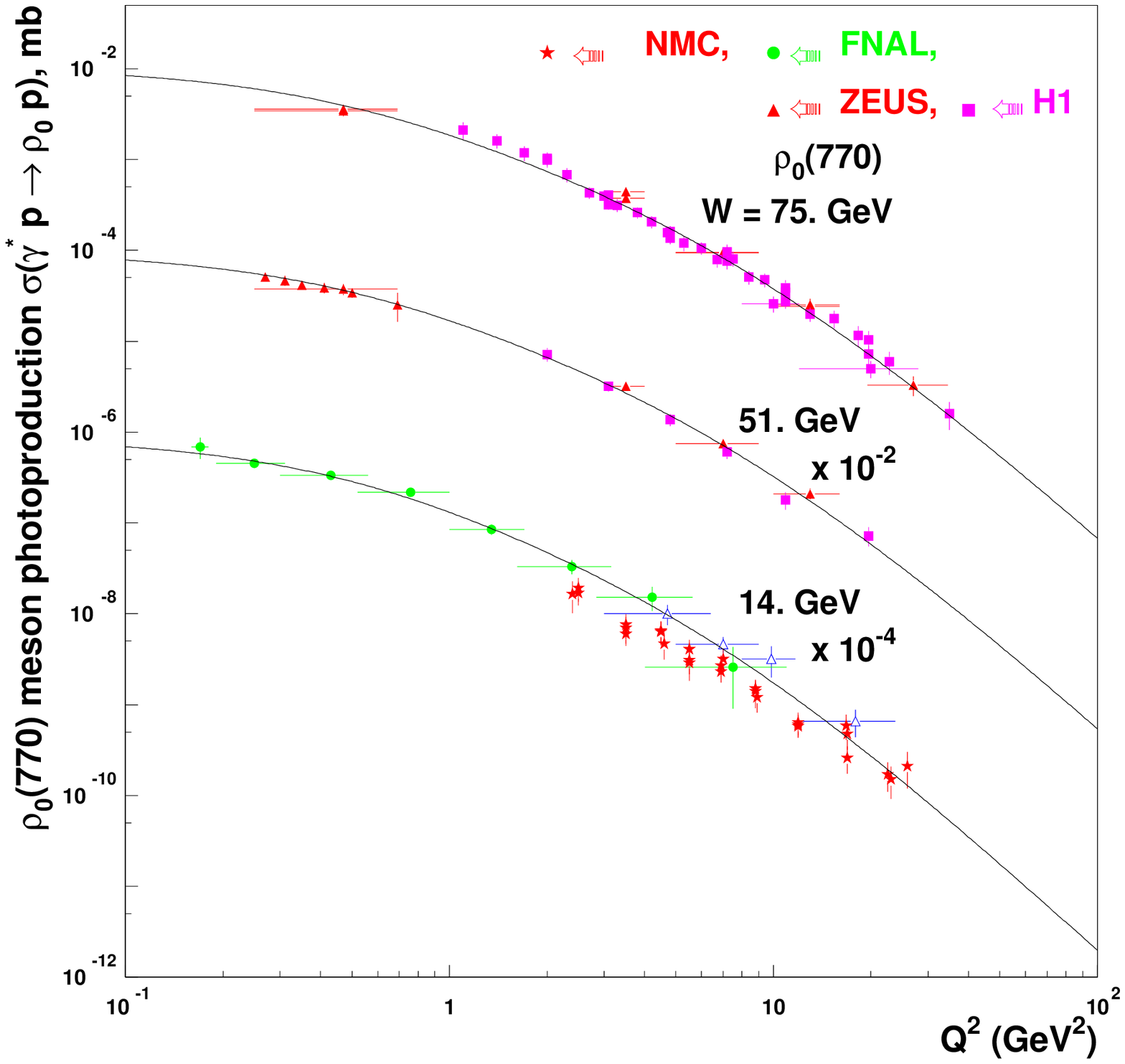}}
\vskip -3.cm
\caption{Elastic cross section of exclusive $\rho_0$ meson photoproduction
by virtual photons as a function of $Q^2$ for
$W=75,\;51,\;{\rm and}\;14\; GeV$. The data and curves for
$W=51,\;{\rm and}\;14\; GeV$ are scaled by factors $10^{-2}$ and $10^{-4}$.
\label{fig:rhoq}
}
\end{figure}

\begin{figure}[H]
\parbox[c]{8.6cm}{\epsfxsize=76mm
\epsffile{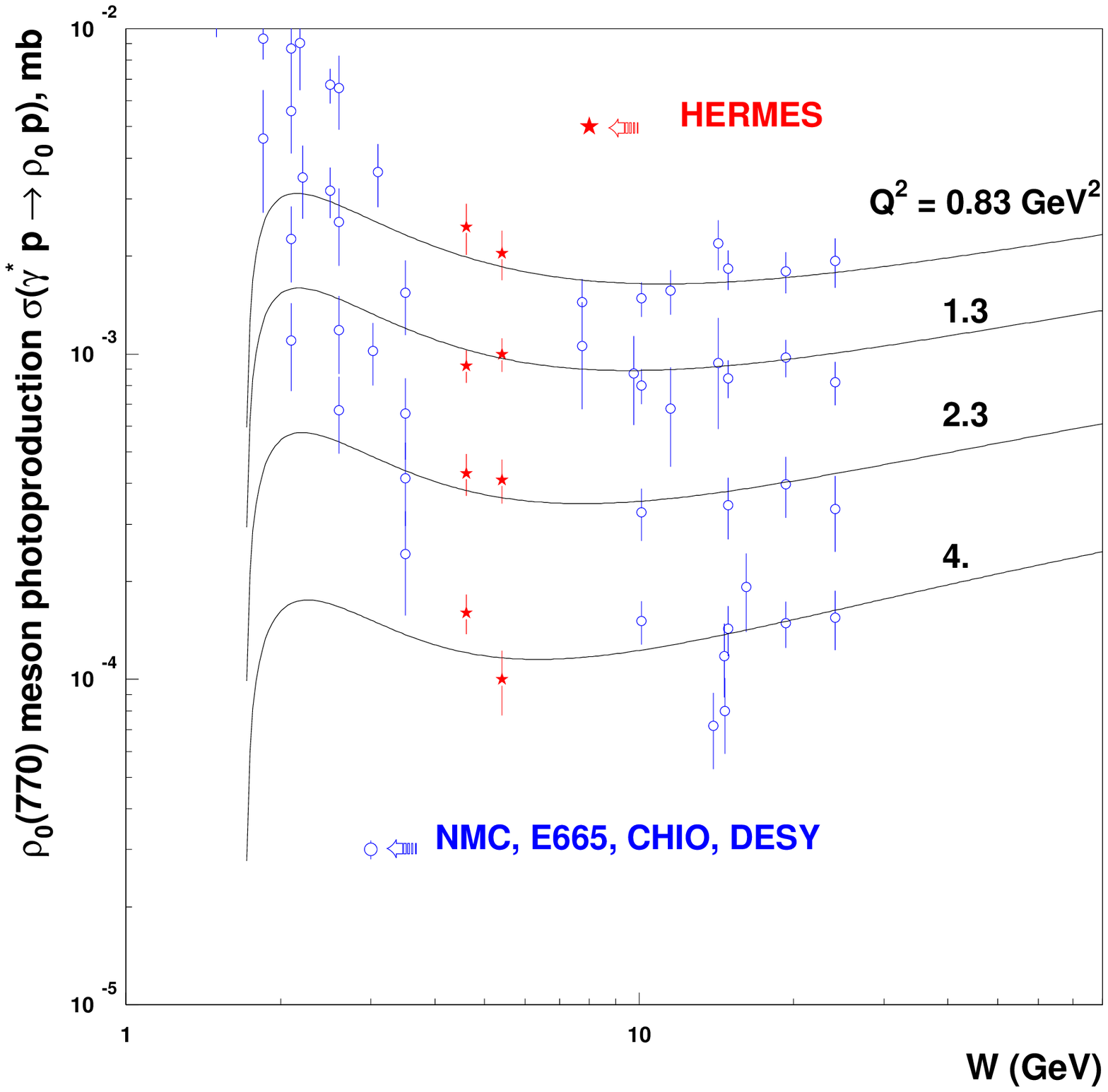}} \hfill~\parbox[c]{7.6cm}{\epsfxsize=76mm
\epsffile{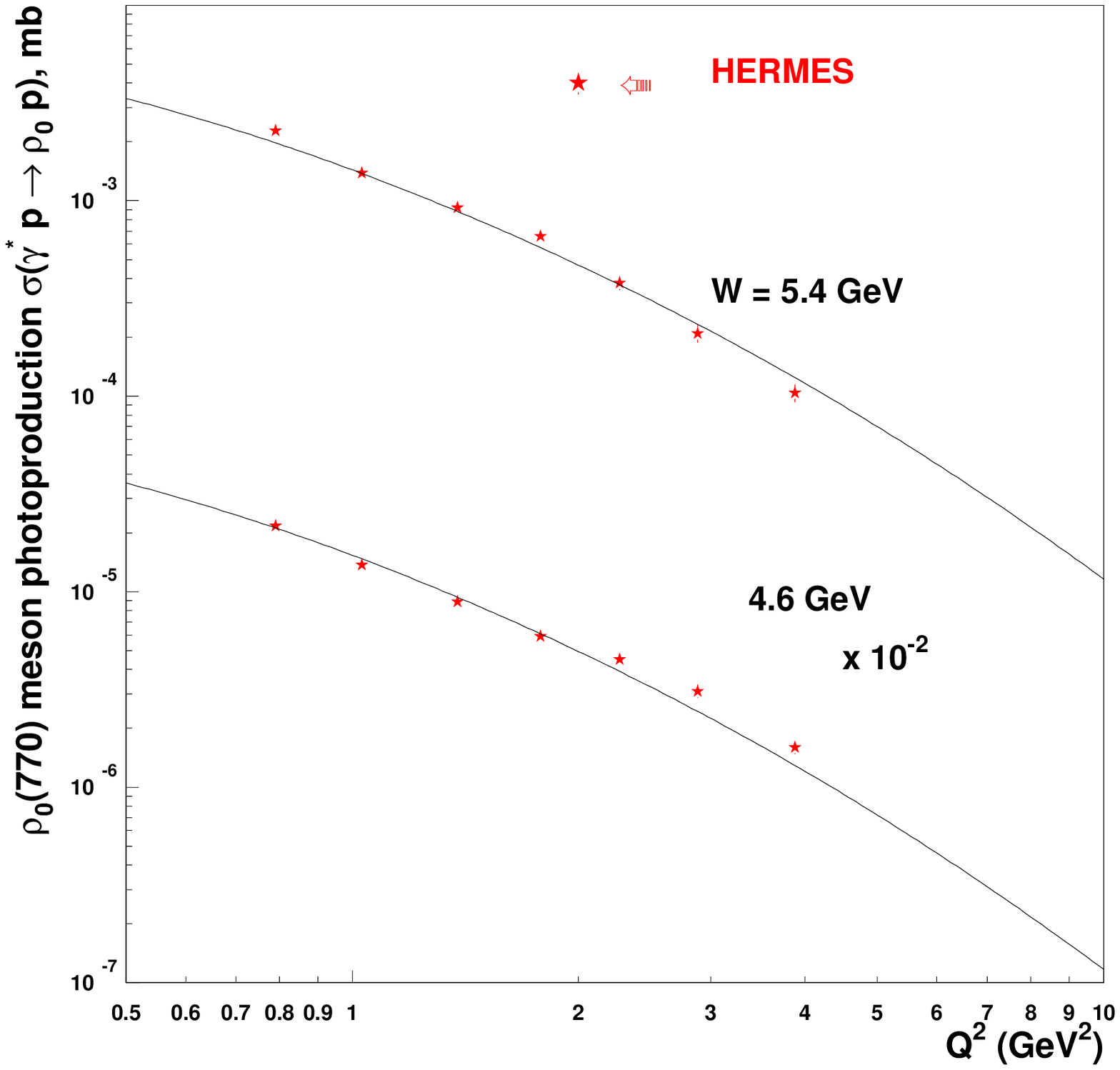}}

\vspace*{-1.3cm}
\parbox[t]{7.7cm}{\caption{Elastic cross section of exclusive $\rho_0$
meson photoproduction by virtual photons as a function of $W$ for
various $Q^2$ in the region of low and intermediate $W$.
\label{fig:rhohermes}}}
\hfill~\parbox[t]{7.7cm}{\caption{Elastic cross section of exclusive
$\rho_0$ meson photoproduction by virtual photons as a function of $Q^2$
for $W=5.4,\;{\rm and}\;4.6\; GeV$. The data and curves for $W=4.6\; GeV$
are scaled by a factor $10^{-2}$.\label{fig:rhoq1}}}
\end{figure}

The results of the fit are depicted in Figs. \ref{fig:rho},
\ref{fig:rhoq}, \ref{fig:rhohermes}, \ref{fig:rhoq1}. The description of
the data is very good in all the range of energies. Both high energy data
from ZEUS and H1 Fig. \ref{fig:rho} and low energy data from HERMES Fig.
\ref{fig:rhohermes} are described. In the region of the HERMES data
(Fig.~\ref{fig:rhohermes}) our description is comparable to the one of
Haakman, Kaidalov and Koch \cite{ref:Kaidalov} (see \cite{ref:Hermes} for
details).



We can now check the predictions of the model. As stated earlier, we aimed
at proposing a unified model for all vector meson production, thus the
only variable that changes is the mass of the vector meson. In the
following figures we depict our {\it predictions} for $\omega$, $\varphi$
and $J/\psi$ mesons and we compare them with the available data. The {\it
description} of the data is very good for all the three mesons. The
$\chi^2=0.83$ for $J/\psi$ meson exclusive production follows without any
fitting. Both $W$ and $Q^2$ dependences are reproduced very well.

\begin{figure}[H]
\parbox[c]{8.7cm}{\epsfxsize=76mm
\epsffile{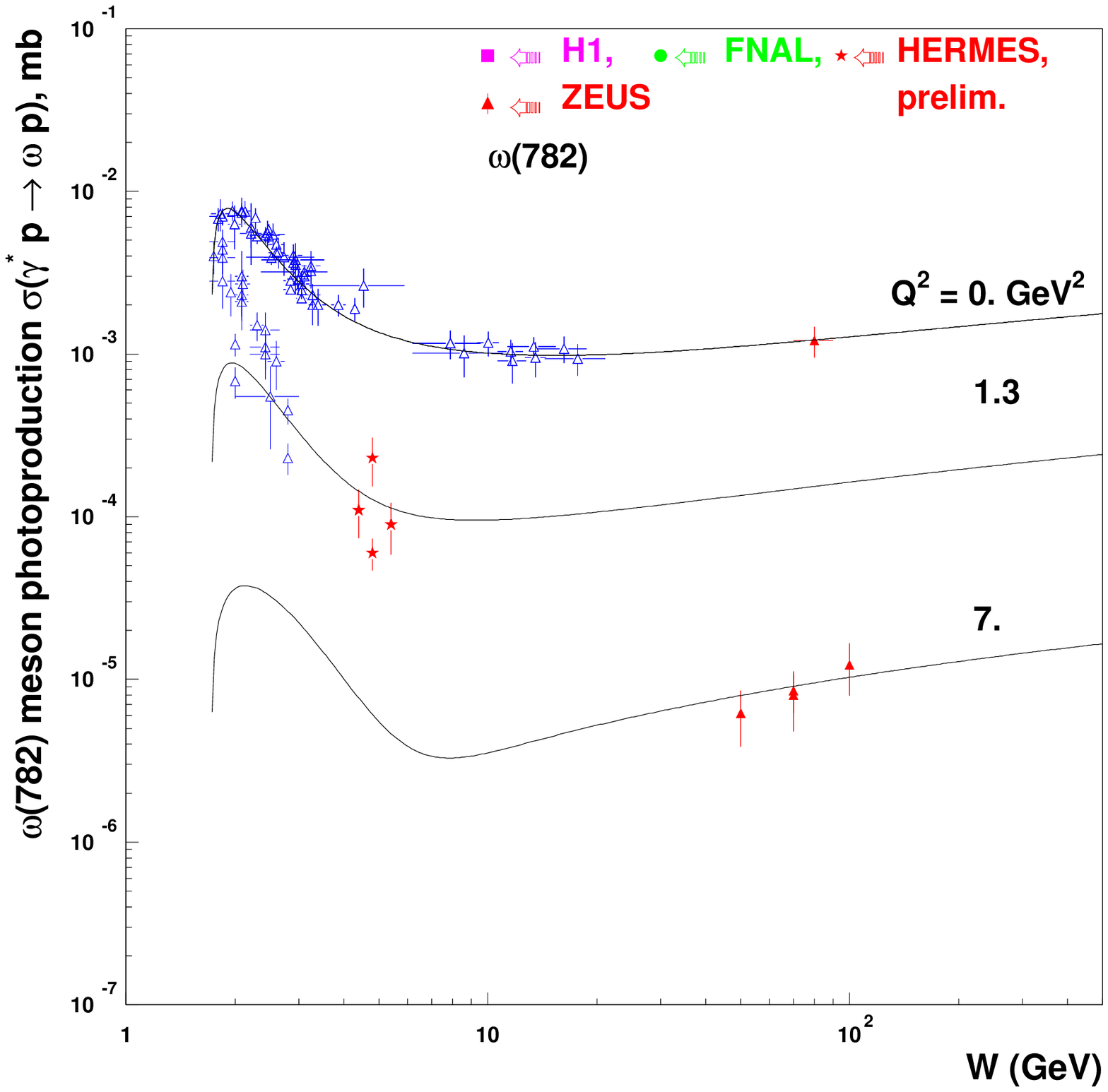}} \hfill~\parbox[c]{7.6cm}{\epsfxsize=76mm
\epsffile{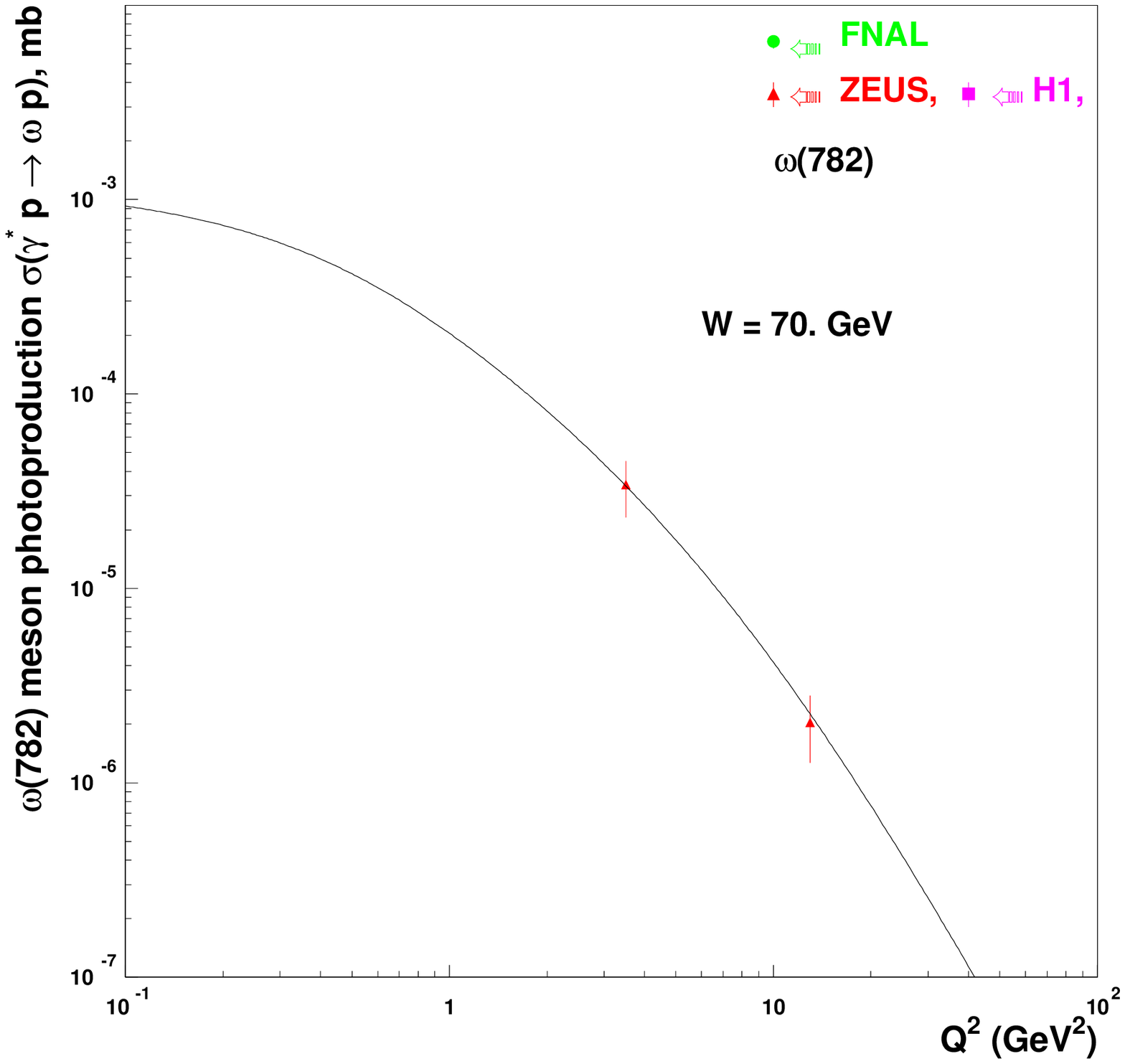}}

\vspace*{-1.6cm}
\parbox[t]{7.7cm}{\caption{Elastic cross section of exclusive $\omega$
meson photoproduction by virtual photons as a function of $W$ for various
$Q^2$.\label{fig:omega}}}
\hfill~\parbox[t]{7.7cm}{\caption{Elastic cross section of exclusive
$\omega$ meson photoproduction by virtual photons as a function of $Q^2$
for $W=70\; GeV$.\label{fig:omegaq70}}}

\vspace*{-0.7cm}

\parbox[c]{8.7cm}{\epsfxsize=76mm
\epsffile{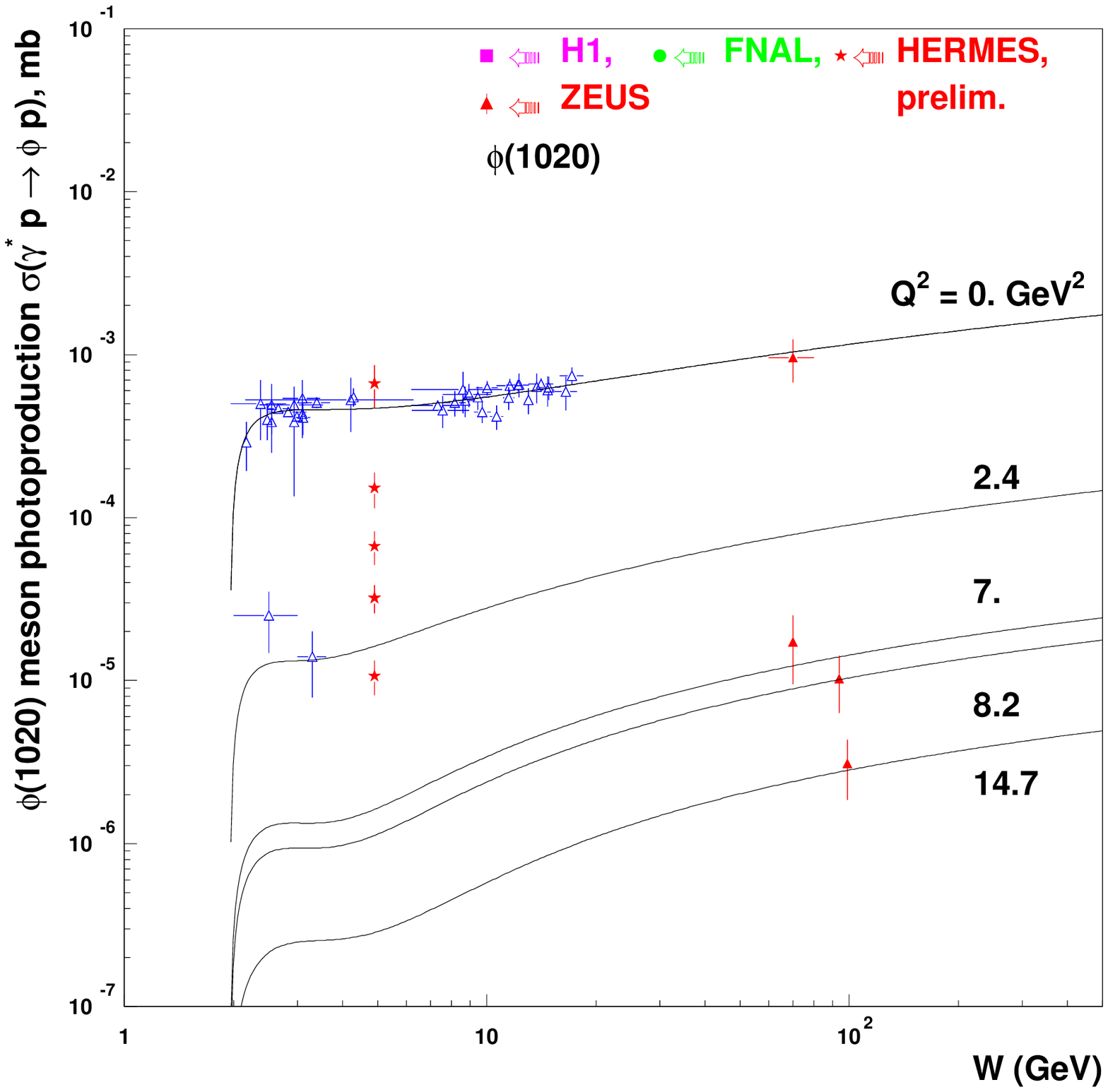}} \hfill~\parbox[c]{7.6cm}{\epsfxsize=76mm
\epsffile{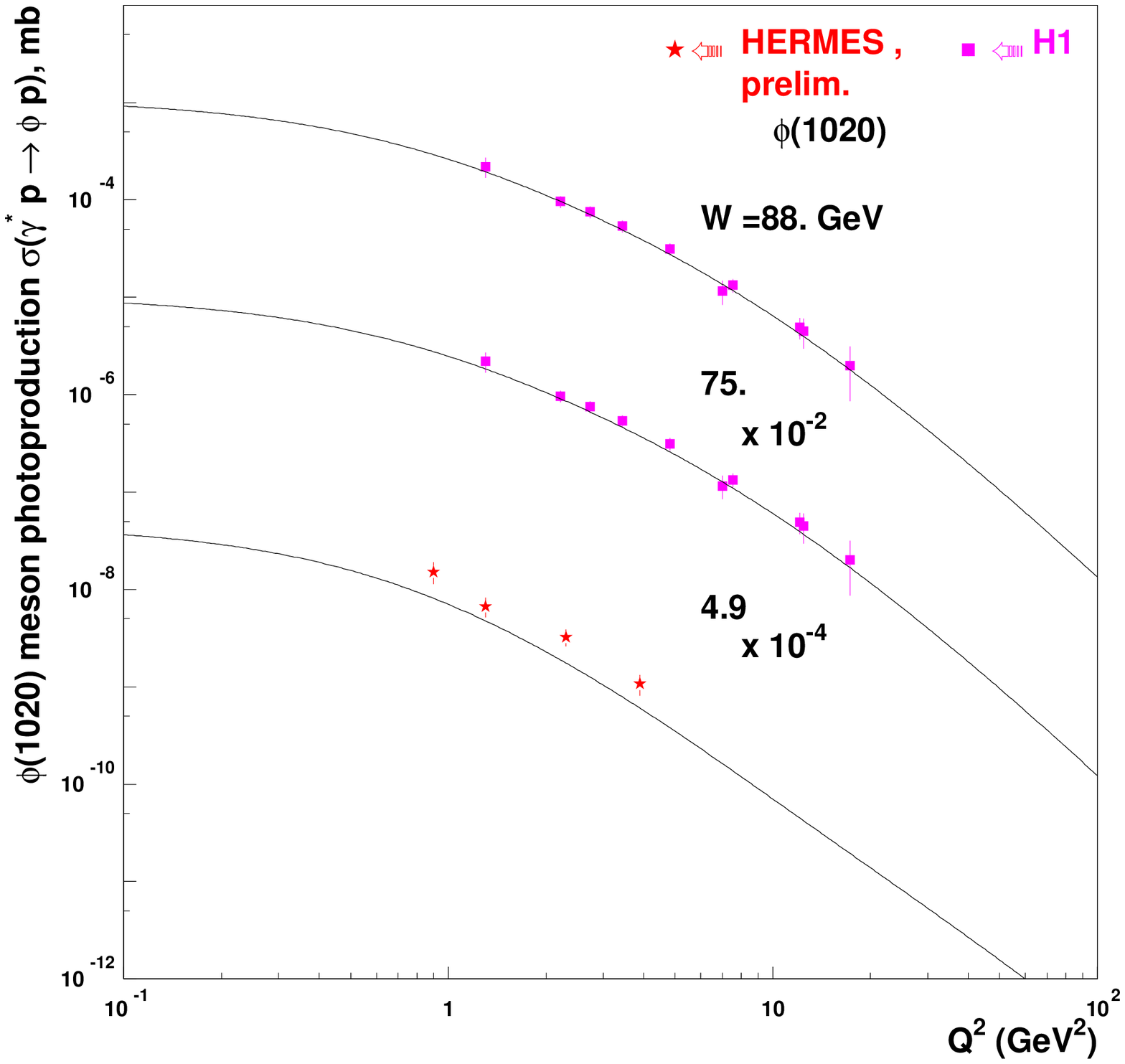}}

\vspace*{-1.6cm}
\parbox[t]{7.7cm}{\caption{Elastic cross section of exclusive $\varphi$
meson photoproduction by virtual photons as a function of $W$ for various
$Q^2$.\label{fig:phi}}}
\hfill~\parbox[t]{7.7cm}{\caption{Elastic cross section of exclusive
$\varphi$ meson photoproduction by virtual photons as a function of $Q^2$
for $W=88, \;75,\; {\rm and}\;4.9\;GeV$  (with scaling factors of
$10^{-2}$ and $10^{-4}$.)
\label{fig:phiq75}}}
\end{figure}

\begin{figure}[H]
\parbox[c]{8.7cm}{\epsfxsize=76mm
\epsffile{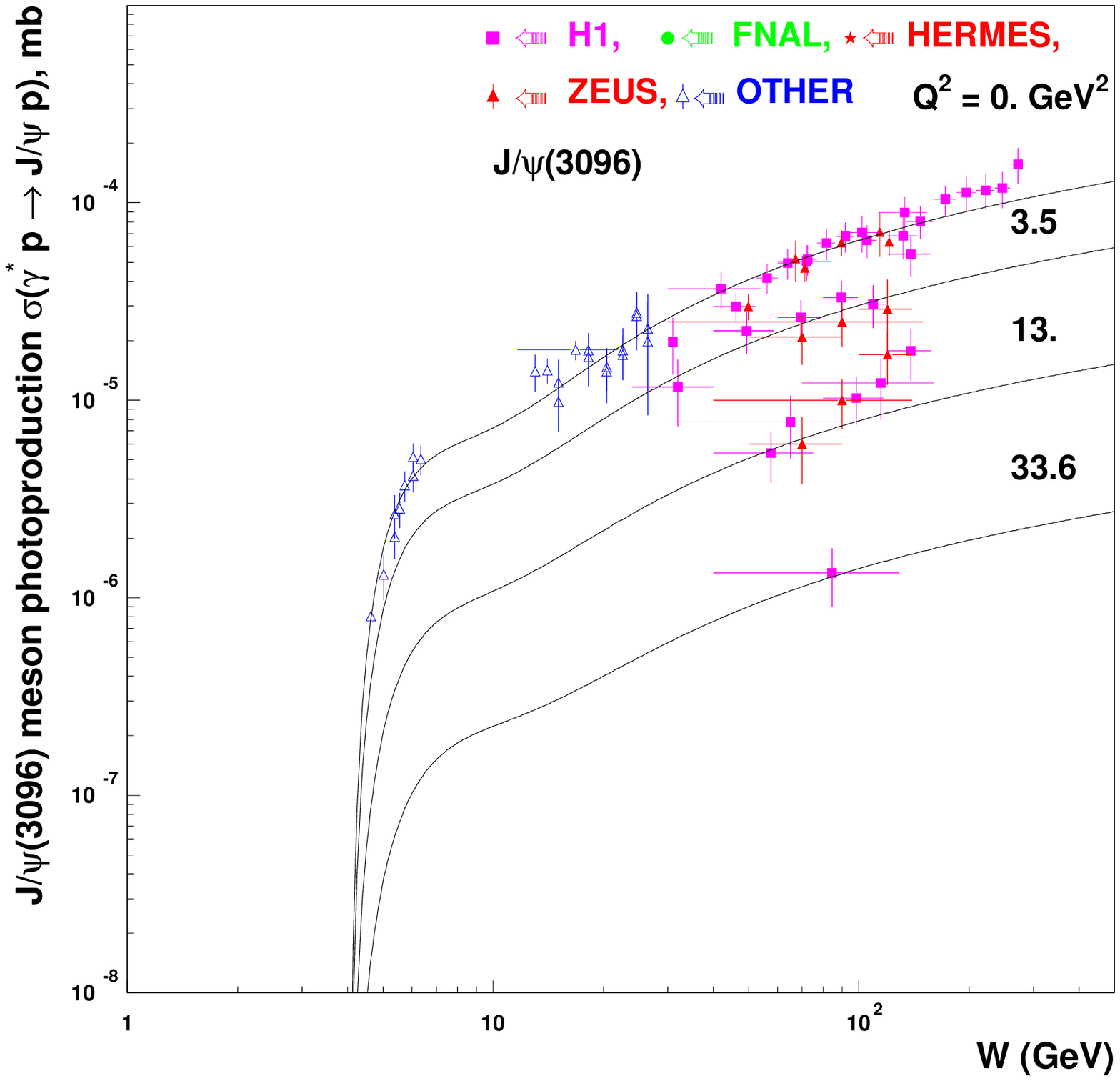}} \hfill~\parbox[c]{7.6cm}{\epsfxsize=76mm
\epsffile{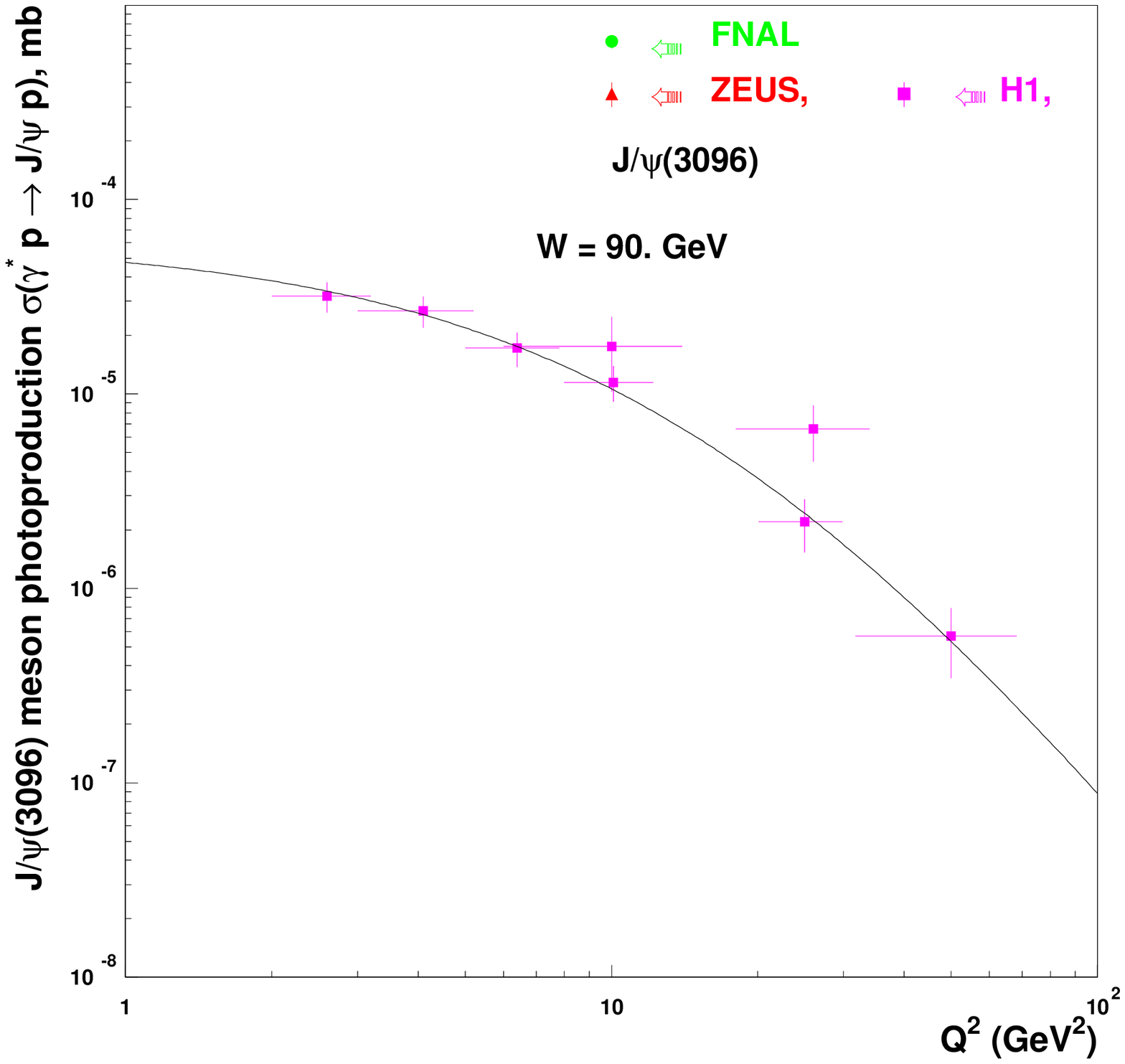}}

\vspace*{-1.3cm}
\parbox[t]{7.7cm}{\caption{Elastic cross section of exclusive $J/\psi$
meson photoproduction by virtual photons as a function of $W$ for various
$Q^2$.\label{fig:jpsi}}}
\hfill~\parbox[t]{7.7cm}{\caption{Elastic cross section of exclusive
$J/\psi$ meson photoproduction by virtual photons as a function of $Q^2$
for $W=90\; GeV$.\label{fig:jpsiq90}}}
\end{figure}

\begin{figure}[H]
\vspace*{-0.5cm}
\parbox[c]{8.7cm}{\epsfxsize=76mm
\epsffile{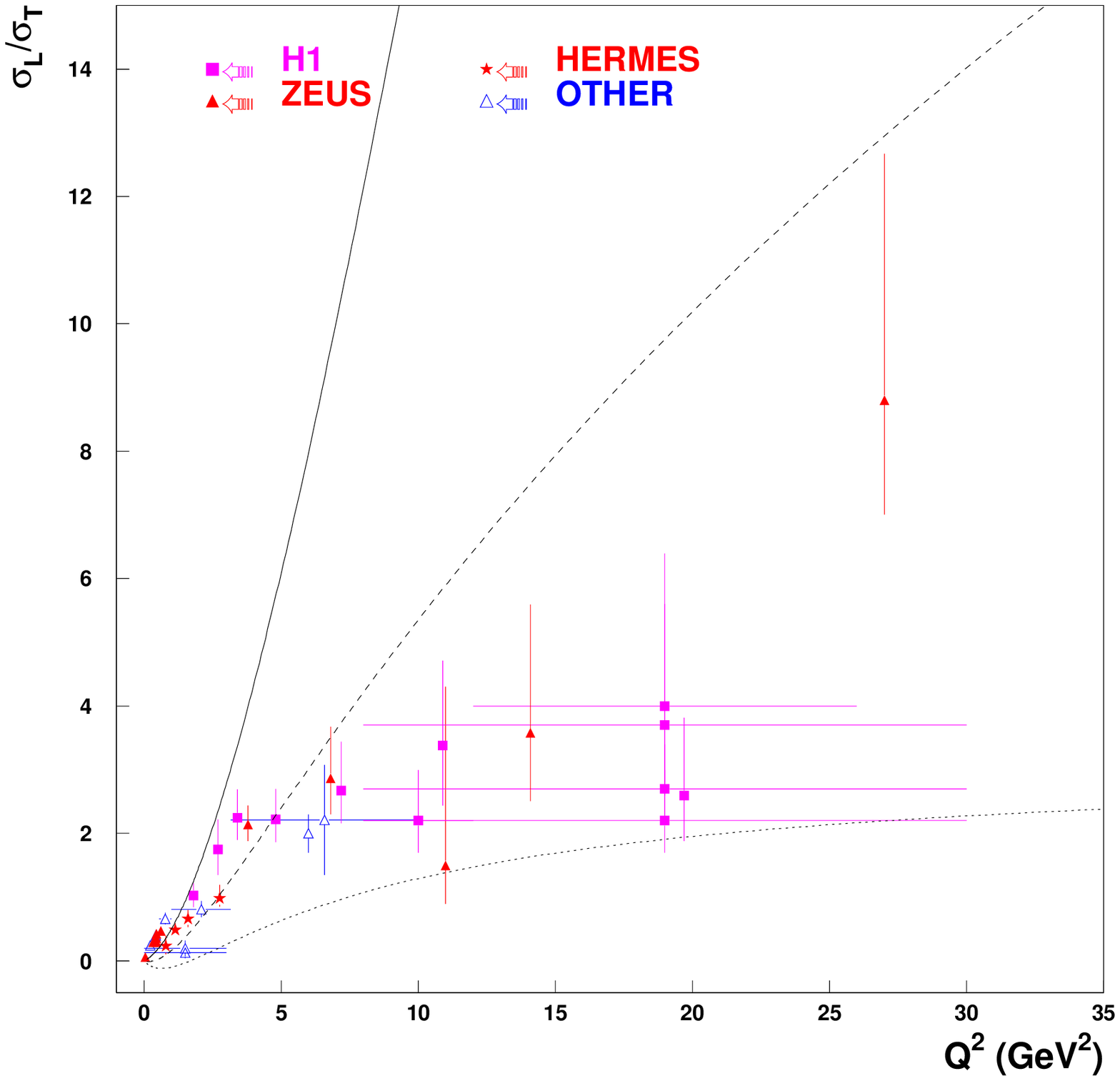}} \hfill~\parbox[c]{7.6cm}{\epsfxsize=76mm
\epsffile{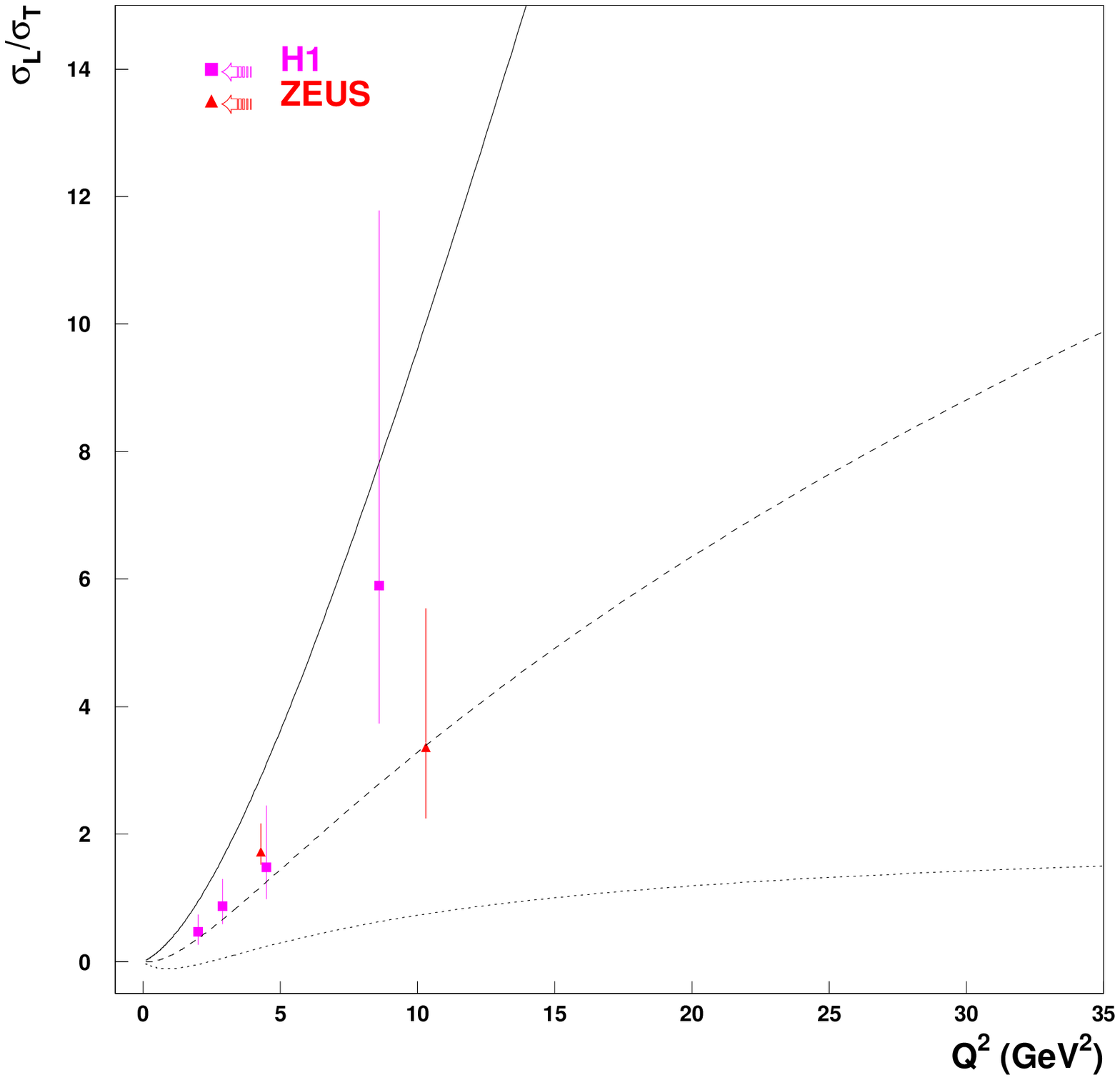}}

\vspace*{-1.7cm}
\parbox[t]{7.7cm}{\caption{Ratio of $\sigma_L/\sigma_T$ for exclusive
$\rho_0$ meson photoproduction.\label{fig:rholt}}}
\hfill~\parbox[t]{7.7cm}{\caption{Ratio of $\sigma_L/\sigma_T$ for
exclusive $\varphi$ meson photoproduction.\label{fig:philt}}}

\vspace*{-.7cm}

\parbox[c]{8.7cm}{\epsfxsize=80mm
\epsffile{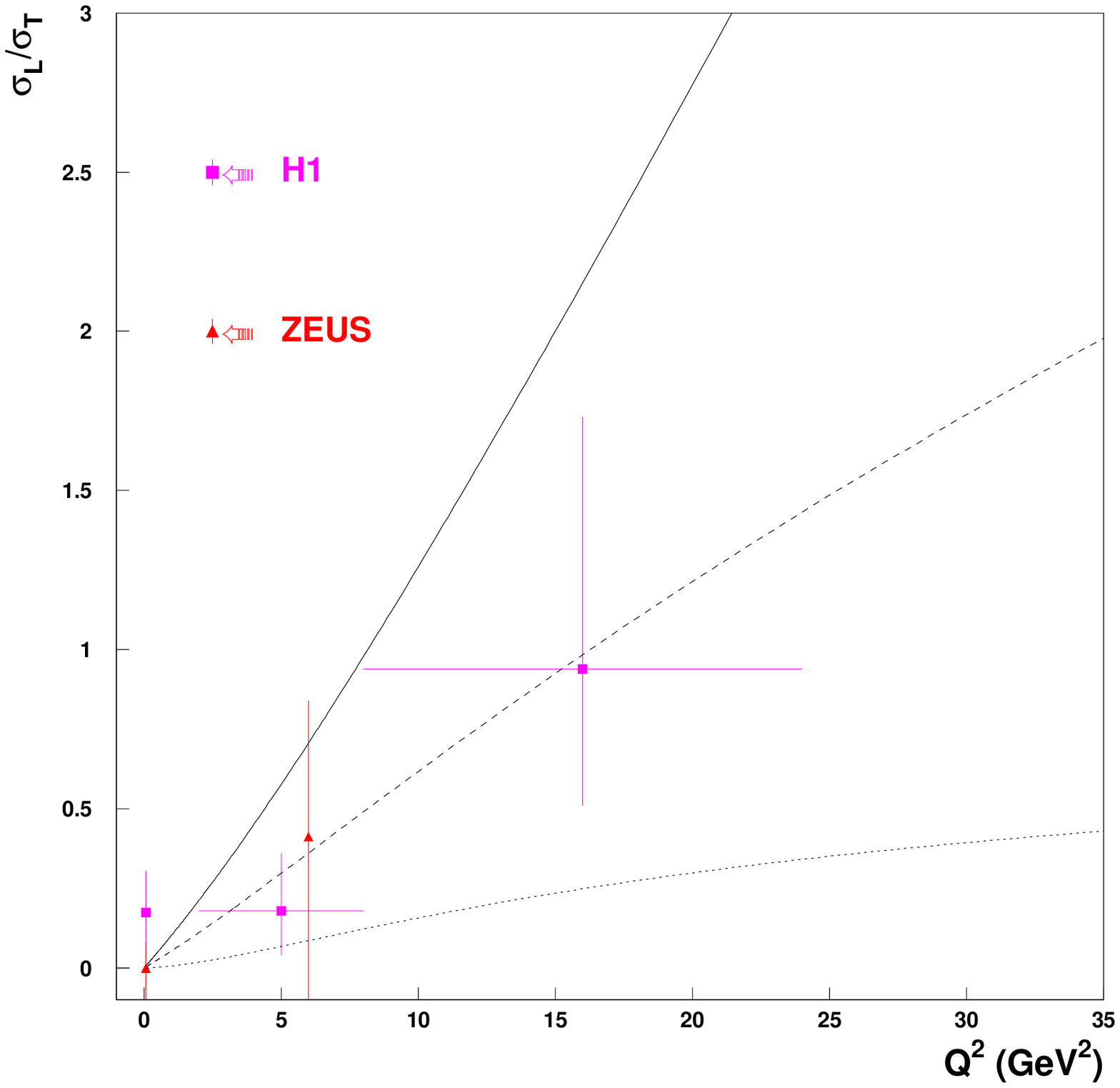}}

\vspace*{-1.7cm}
\parbox[t]{7.7cm}{\caption{Ratio of $\sigma_L/\sigma_T$ for exclusive
$J/\psi$ meson photoproduction.\label{fig:jpsilt}}}

\end{figure}

We can also plot the various ratios $\sigma_L/\sigma_T$ ( shown in Fig.
\ref{fig:rholt} and \ref{fig:philt}, \ref{fig:jpsilt}) which , we recall,
are constrained by our ``ad hoc'' choice that the large $Q^2$ QCD
predictions Eq. \ref{eq:qcd} be satisfied. The result shows, indeed, a
rapid increase of $\sigma_L/\sigma_T$ with increasing $Q^2$ but this, at
least for exclusive $\rho_0$ meson virtual photoproduction is not in
agreement with the data (see Fig. \ref{fig:rholt}). This leaves open the
question already discussed in \cite{ref:Cudell1} of modifying
appropriately the QCD predictions Eq. \ref{eq:qcd} to account for the
data. This could be easily done within our model by reconsidering the
choice of introducing the factor $f(Q^2)$ in Eq.~\ref{eq:couplingsq}.

Indeed, we can write \be \sigma = (1+R)\sigma_T
=(1+R)\frac{f(Q^2)}{f(Q^2)}\sigma_T=(1+\tilde R) \tilde \sigma_T\; , \ee
where $\tilde R = (1+R)/f(Q^2)-1$, $\tilde \sigma_T = f(Q^2)\sigma_T$. By
setting $f(Q^2)=\Big(\frac{W_0^2+Q^2+M_V^2}{W_0^2+M_V^2}\Big)^a$ (see Eq.~
\ref{eq:couplingsq}). If we choose $a=0,0.5,\; {\rm or}\; 1$, we can
obtain various asymptotical behaviors for $\sigma_T$ \bea \nonumber
a=0,\;\;\tilde \sigma_T|_{Q^2\rightarrow \infty} \propto \frac{1}{Q^8}\; ; \\
\nonumber
a=0.5,\;\;\tilde \sigma_T|_{Q^2\rightarrow \infty} \propto \frac{1}{Q^7}\; ; \\
\nonumber
a=1,\;\;\tilde \sigma_T|_{Q^2\rightarrow \infty} \propto \frac{1}{Q^6}\; .
\eea

In the figures \ref{fig:rholt}, \ref{fig:philt}, \ref{fig:jpsilt} we show
possible adjustments of $Q^2$ behaviour of $\tilde\sigma_L/\tilde\sigma_T$. Solid line
corresponds to $\tilde\sigma_T\propto\frac{1}{Q^8}$, dashed line corresponds to
$\tilde\sigma_T\propto\frac{1}{Q^7}$ and dotted line corresponds to
$\tilde\sigma_T\propto\frac{1}{Q^6}$.

From this pure phenomenological analysis we would say that the data prefer
$a=0.5$, though we believe that more complete and accurate set of data is
needed in order to resolve the question on the $Q^2$ dependence of the
ratio.

\subsection{Differential cross section of vector meson exclusive production}

We have constructed the amplitude $A(W^2,t;Q^2, M_V^2)$ and made sure that
the model is valid for integrated observable such as elastic cross
section. The natural question of applicability of such an amplitude to
describe also angular distributions arises.

The differential cross section is the following:
\be
\frac{d\sigma}{dt} = 4\pi |A(W,t;Q^2, M_V^2)|^2
\ee

\begin{figure}[H]
\parbox[c]{8.7cm}{\epsfxsize=76mm
\epsffile{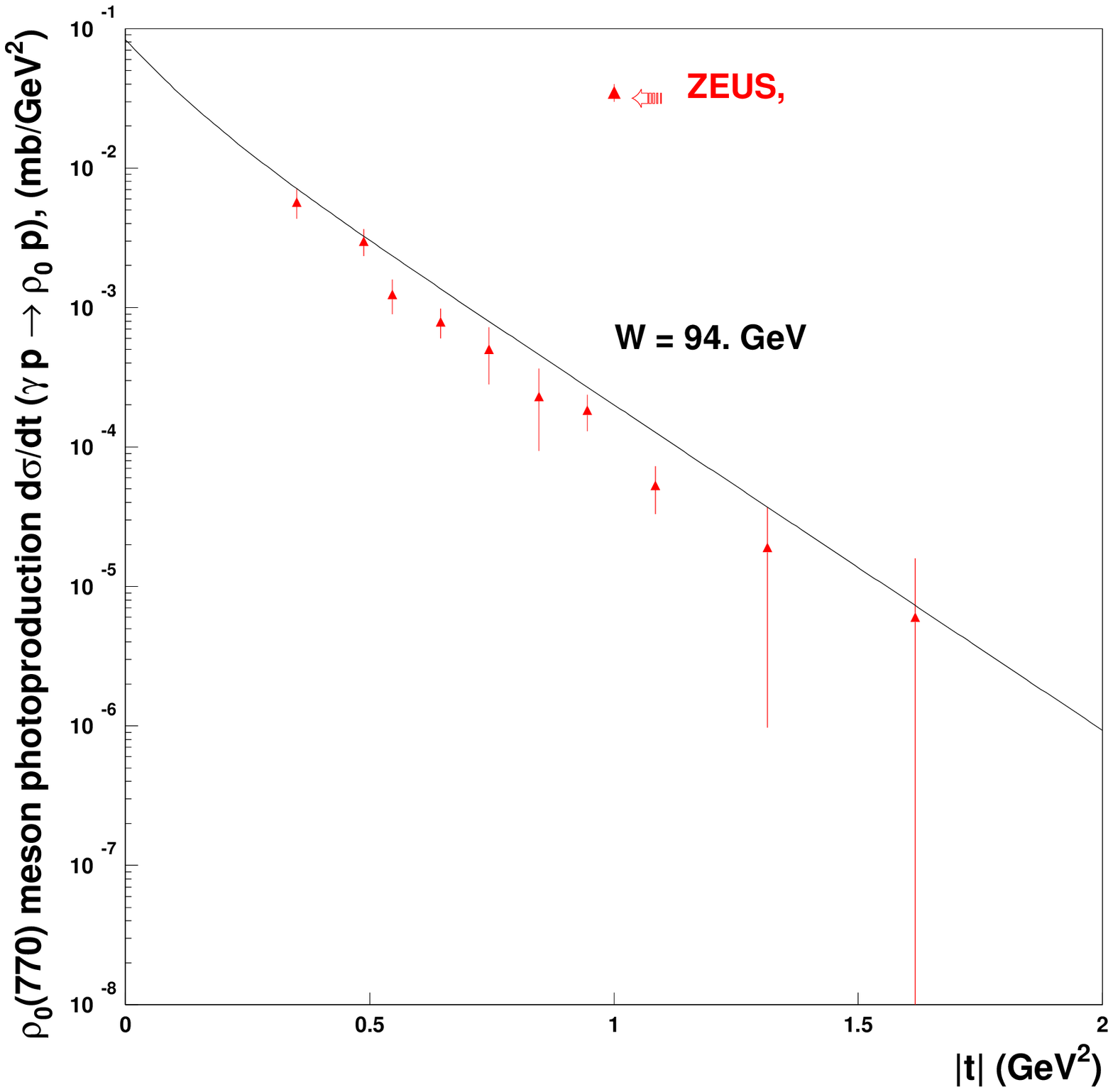}} \hfill~\parbox[c]{7.6cm}{\epsfxsize=76mm
\epsffile{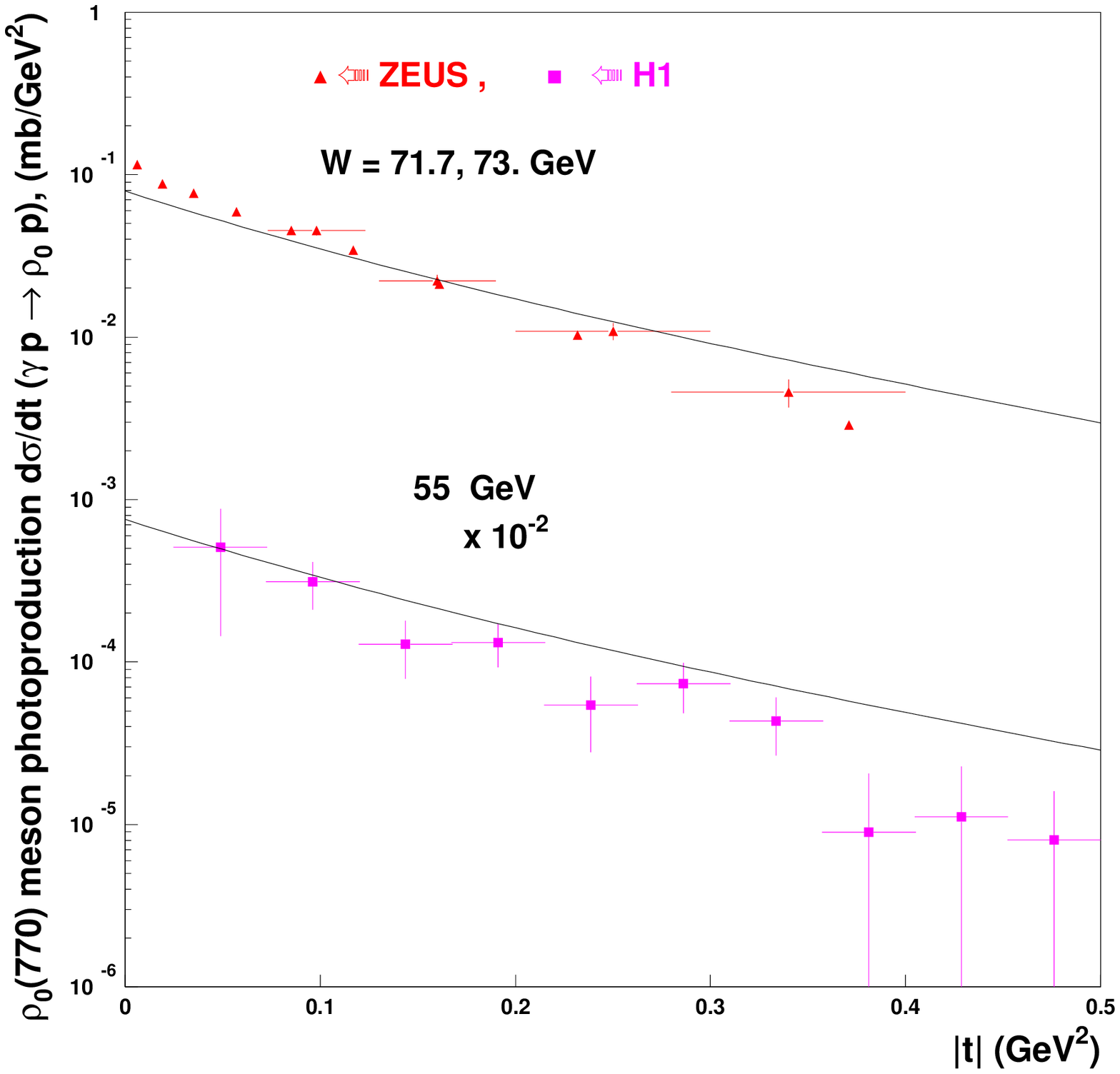}}

\vspace*{-1.3cm}
\parbox[t]{7.7cm}{\caption{Differential cross section of exclusive
$\rho_0$ meson photoproduction for $W=94\; GeV$.\label{fig:rhod}}}
\hfill~\parbox[t]{7.7cm}{\caption{Differential cross section of exclusive
$\rho_0$ meson photoproduction for $W=71.7,\; 73,\; {\rm and}\; 55\; GeV$.
The data and curves for $W=55\; GeV$ are scaled by a factor $10^{-2}$.
\label{fig:rhod1}}}
\end{figure}

\begin{figure}[H]
\centering {\vspace*{ -1cm} \epsfxsize=120mm \epsffile{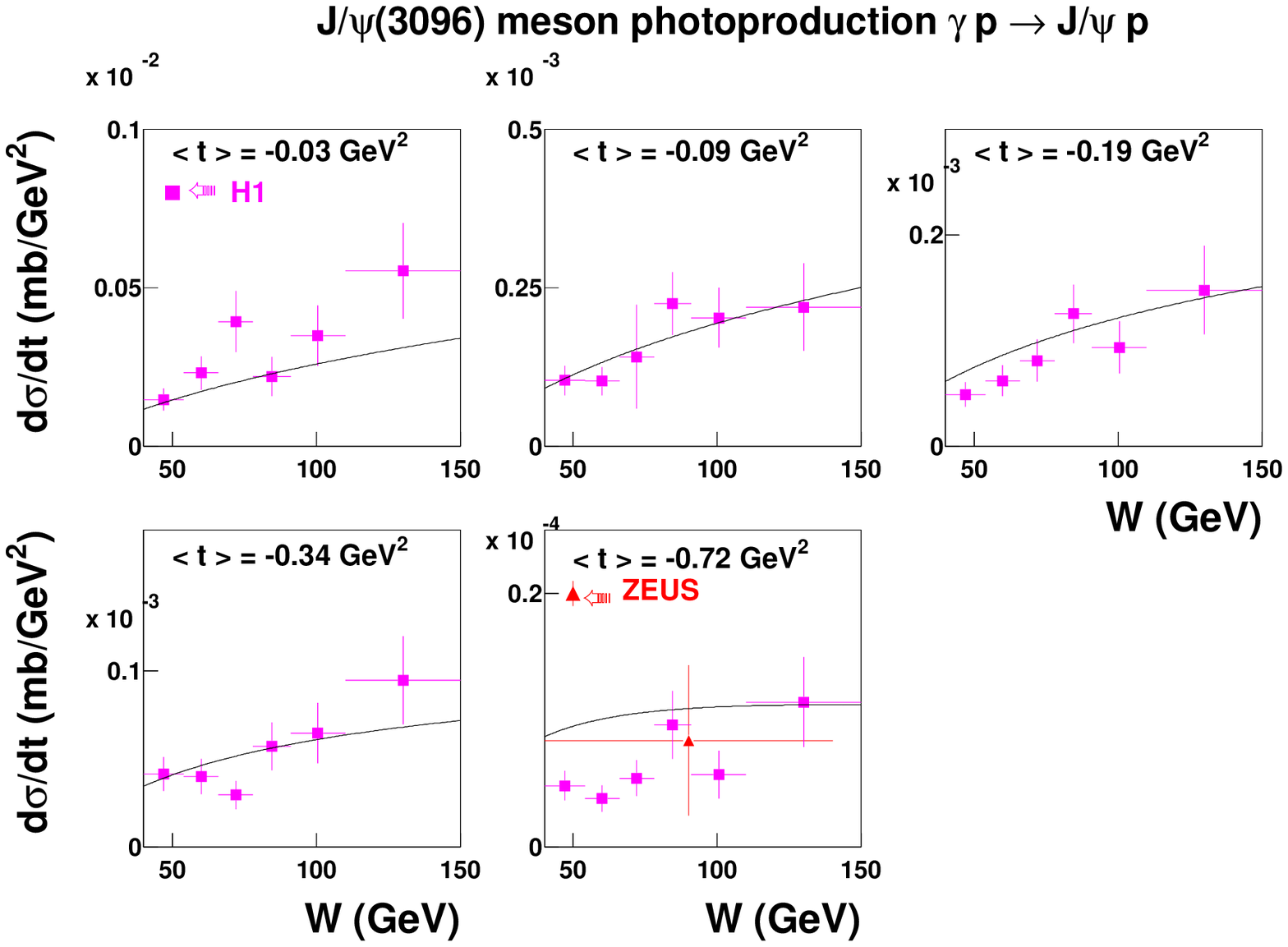}} \vskip
-6.7cm \caption{Differential cross section of exclusive $J/\psi$ meson
photoproduction as a function of $W$ at different $<t>$.
\label{fig:jpsidt} }
\end{figure}

\begin{figure}[H]
\parbox[c]{8.7cm}{\epsfxsize=76mm
\epsffile{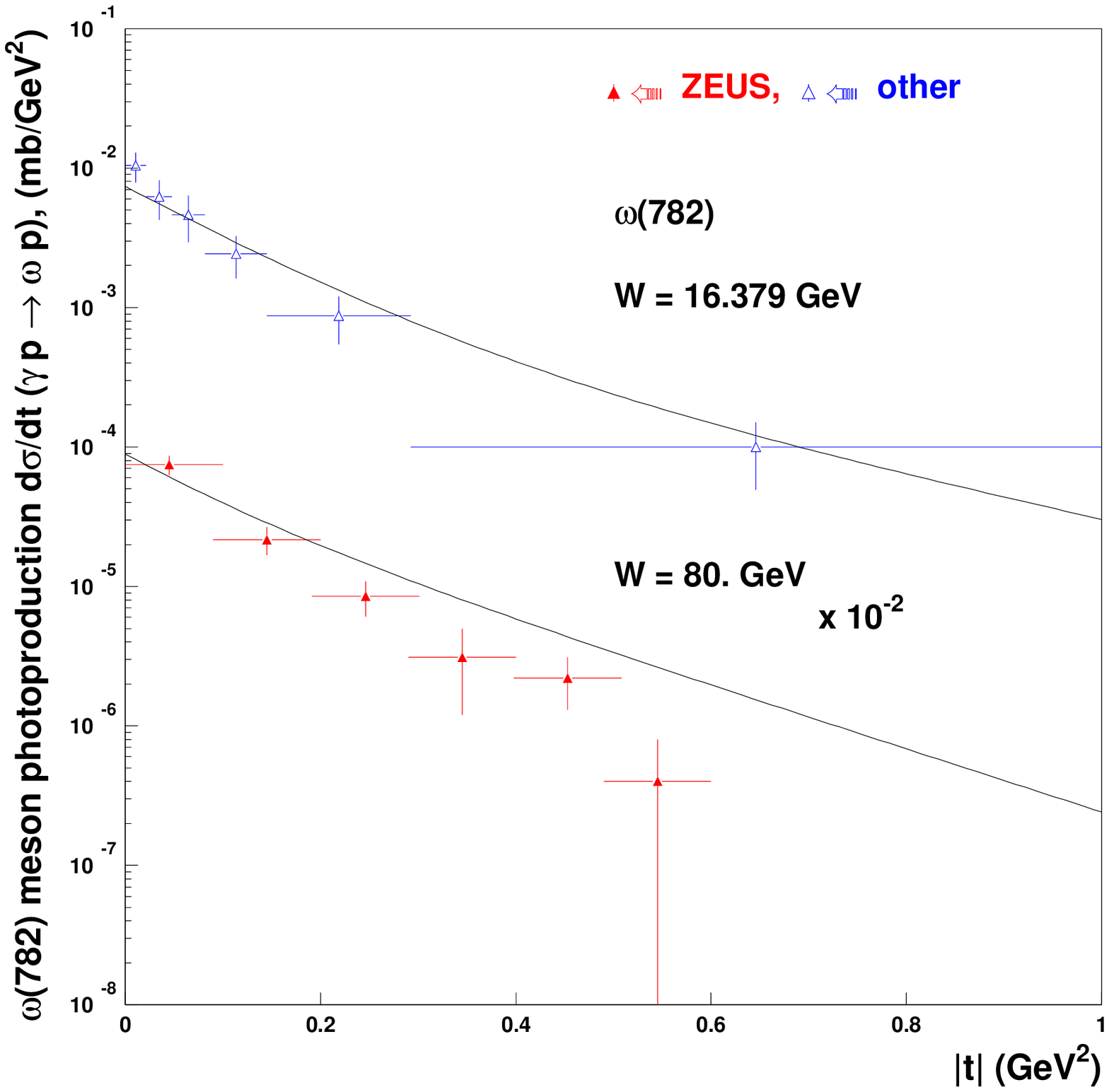}} \hfill~\parbox[c]{7.6cm}{\epsfxsize=76mm
\epsffile{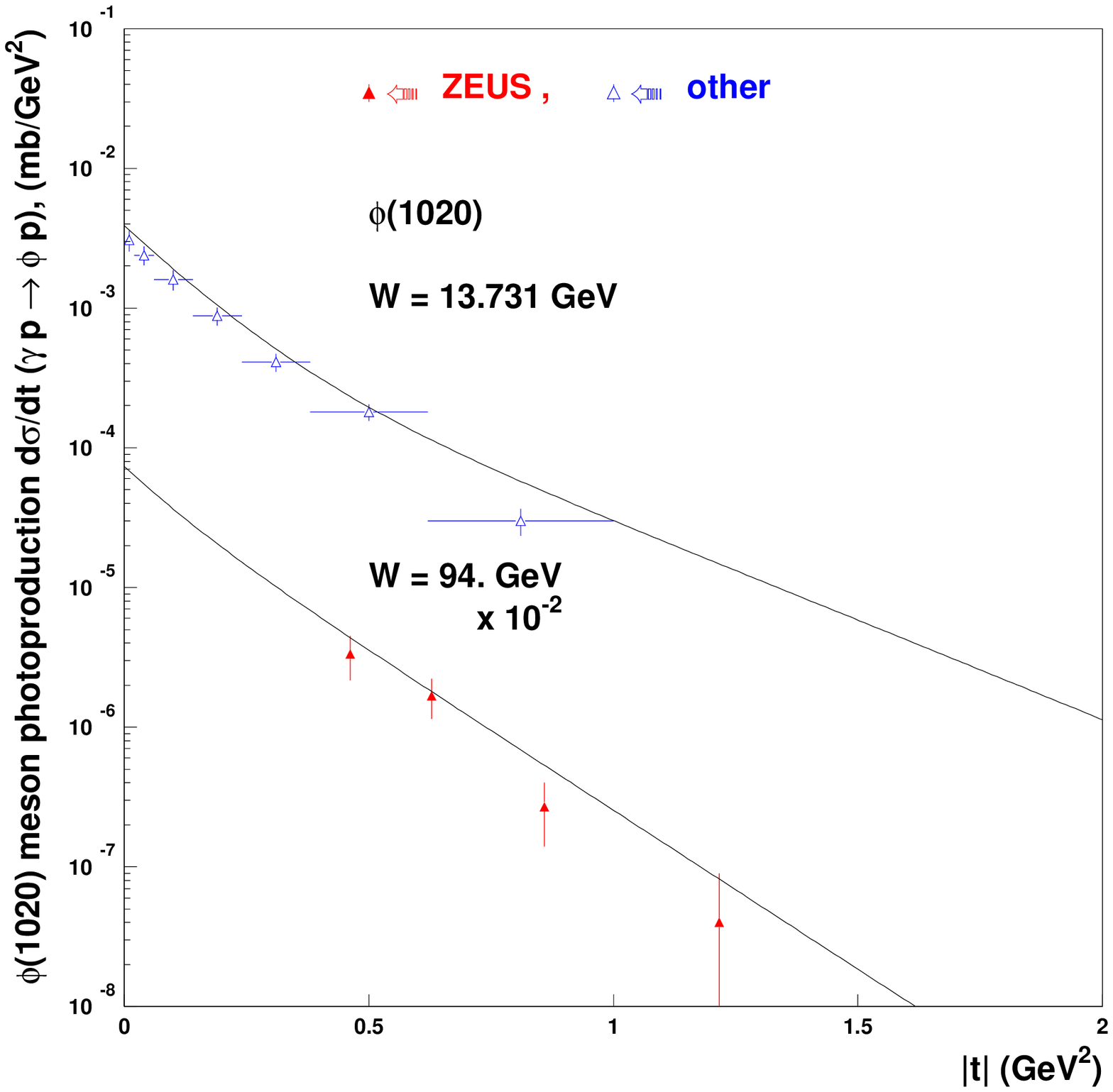}}

\vspace*{-1.7cm}
\parbox[t]{7.7cm}{\caption{Differential cross section of exclusive
$\omega$ meson photoproduction for $W=16.379\;{\rm and}\;80\; GeV$. The
data and curves for $W=80\; GeV$ are scaled by a factor $10^{-2}$.
\label{fig:omegad}}}
\hfill~\parbox[t]{7.7cm}{\caption{Differential cross section of
exclusive $\varphi$ meson photoproduction for $W=13.371\;{\rm and}\; 94\;
GeV$. The data and curves for $W=94\; GeV$ are scaled by a factor $10^{-2}$.
\label{fig:phid}}}
\end{figure}

\begin{figure}[H]
\parbox[c]{8.7cm}{\epsfxsize=76mm
\epsffile{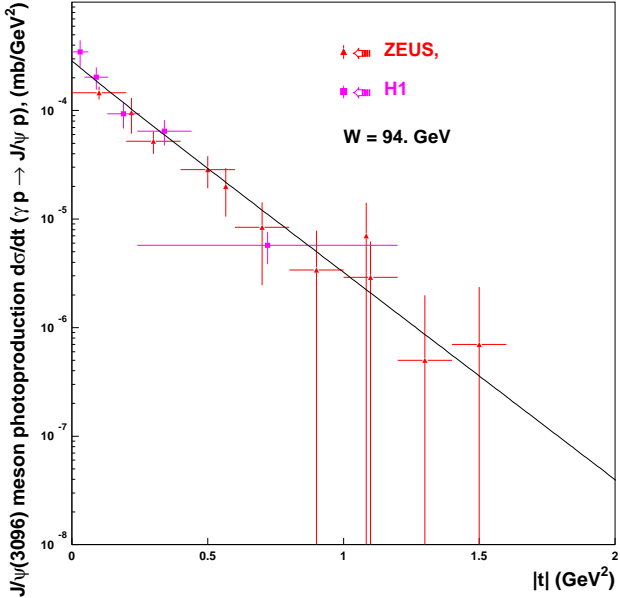}}
\vspace*{-1.3cm}
\parbox[t]{7.cm}{\caption{Differential cross section of exclusive
$J/\psi$ meson photoproduction for $W=94. \; GeV$.\label{fig:jpsid}}}
\end{figure}

Using the amplitude from the previous section this quantity can be
calculated and the comparison with the data is presented in Fig.
\ref{fig:rhod}, \ref{fig:rhod1}, \ref{fig:jpsid}, \ref{fig:jpsidt},
\ref{fig:omegad} and \ref{fig:phid}.

The description is very good without any additional fitting and we find
that $\alpha'_\Pom(0)= 0.25 \;(GeV^{-2})$ is universal for both
hadron-hadron scattering and exclusive vector meson photoproduction.

In view of our result concerning the universality of our approach we are
led to conclude that extracting the Pomeron trajectory from the
experimental data as proposed in \cite{ref:H1JPSI} using the data depicted
in Fig.~\ref{fig:jpsidt} cannot be regarded as a valid support to the need
of either hard Pomeron contribution or NLO BFKL Pomeron.

\vspace*{ -1cm}


\section{Conclusion}

We have shown that all available data on photoproduction of the
various vector mesons can be described in the framework of the Regge
approach. Not only the cross sections at $Q^{2}=0$ but also the data at
$Q^{2}\neq 0$ are described with a good quality. Moreover our model
describes without additional fitting (as far as all parameter are
determined from the fit to $t=0$ data) also the differential
cross-sections of vector meson production. We would like to emphasize the
following important points
\begin{enumerate}
\item
Pomeron and secondary Reggeons appear as universal objects in Regge
theory. So the corresponding $j$-singularities of $\gamma$ p amplitudes
and their trajectories (intercepts $\alpha(0)$ and slopes $\alpha'$ in a
linear approximation) at $Q^{2}=0$ coincide with those in pure hadronic
amplitudes. They do not depend on the properties of external particles
and, consequently, on $Q^{2}$. The unitarity restrictions on Pomeron
contribution obtained strictly for the $hh$ case has to be valid for
$\gamma h$ one as far as it is universal.
\item
The growth with energy of hadronic total cross sections and the
 restriction on the
Pomeron intercept ($\alpha_{\Pom}(0)\leq 1$) implied by the
Froissart-Martin bound mean that the Pomeron is a more complicated
singularity than a simple pole with $\alpha_{\Pom}(0)=1$. In accordance
with the above mentioned universality this statement can be extended to
the DIS case.

We have considered the simplest case when the Pomeron is a double $j$-pole
leading to $\sigma(s)\propto lns$. Thus double pole (or dipole) Pomeron
model describes well hadronic reactions (total and differential
cross-sections) and the proton structure function.
\item
There are no new $j$-singularities in the DIS amplitudes $Q^{2}\neq 0$.
In particular, no hard Pomeron contribution appears necessary
in the present regime of $Q^{2}$ and $t$.
\item
The model we suggest is quite economic. The only parameter that makes
the transition from one photoproduction
process to another is the mass of
vector meson $m_V$.
\item
Our model describes the data also at
low energies due to the threshold factor. This is particularly
important for $J/\psi$ production where the bulk of available
data are not so far from threshold.
\item
Many simplifying assumptions have been implicitly made, for example, some
parameters were taken the same for different terms of the amplitudes. When
more precise data will become available, some of these simplifying
assumptions need not be valid anylonger. So the model has a potential to
be improved if necessary.

\end{enumerate}
\noindent
{\bf Note added in proof.} After this paper was completed, we learned about a paper
\cite{ref:Jenk2}
by Fiore, Jenkovszky, and Paccanoni where some of the arguments discussed here
are covered.

\subsection*{Acknowledgement}
We would like to thank Alexander Borissov for various and fruitful
discussions.


\end{document}